\documentclass[aps,article,nofootinbib,groupedaddress]{revtex4}
\usepackage{graphicx}
\usepackage{amsmath,amssymb}
\usepackage{hyperref}
\usepackage[usenames]{color}
\usepackage{bm}

\hypersetup{
    colorlinks=true,
    linkcolor=red,
    citecolor=blue,
}

\def\nn{\nonumber}

\newcommand{\be}{\begin{equation}}
\newcommand{\ee}{\end{equation}}
\newcommand{\bq}{\begin{eqnarray}}
\newcommand{\eq}{\end{eqnarray}}
\newcommand{\bsq}{\begin{subequations}}
\newcommand{\esq}{\end{subequations}}
\newcommand{\bc}{\begin{center}}
\newcommand{\ec}{\end{center}}

\newcommand{\gsim}{\raise.3ex\hbox{$>$\kern-.75em\lower1ex\hbox{$\sim$}}}
\newcommand{\lsim}{\raise.3ex\hbox{$<$\kern-.75em\lower1ex\hbox{$\sim$}}}
\newcommand{\cs}{c_{_{\rm S}}}
\newcommand{\fnl}{f_{_{\rm NL}}}

\newcommand{\nc}{\newcommand}

\newcommand{\Mp}{M_\mathrm{Pl}}

\newcommand{\uin}{\mathrm{in}}
\newcommand{\nS}{n_{_{\mathrm S}}}

\newcommand{\ie}{\emph{i.e.~}}

\newcommand{\eg}{\emph{e.g.~}}

\newcommand\bgammap{\gamma _{_{\rm SR}}^{(+)}}
\newcommand\bgammam{\gamma _{_{\rm SR}}^{(-)}}

\newcommand{\CAMB}{{\tt CAMB}}
\newcommand{\MCMC}{{\tt COSMOMC}}

\nc{\ba}{\begin{eqnarray}}
\nc{\ea}{\end{eqnarray}}

\nc{\dd}{{\rm d}}

\def\nn{\nonumber}

\def\fnl{f_{_{\rm NL}}}

\begin{document}

\title{Superimposed Oscillations in Brane Inflation}

\author{Santiago \'Avila\footnote{E-mail:
    santiago.avila@uam.es}$^{a,c}$, J\'er\^ome Martin\footnote{E-mail:
    jmartin@iap.fr}$^{b}$, Dani\`ele A.~Steer\footnote{E-mail:
    steer@apc.univ-paris7.fr}$^{c,b}$}

\affiliation{$^a$ Departamento de F\'isica Te\'orica, Facultad de Ciencias,
Universidad Aut\'onoma de Madrid, 28049 Cantoblanco, Madrid, Spain \\
  $^b$Institut Astrophysique de Paris, UMR 7095-CNRS, Universit\'e Pierre 
et Marie Curie, 98bis Boulevard Arago, 75014 Paris, France\\
  $^c$APC, Universit\'e Paris-Diderot, CNRS/IN2P3, CEA/IRFU
    and Observatoire de Paris, 10 rue Alice Domon et L\'eonie Duquet,
  75205 Paris Cedex 13, France }

\date{\today}

\begin{abstract}
  In canonical scalar field inflation, the Starobinsky model (with a
  linear potential but discontinuous slope) is remarkable in that
  though slow-roll is violated, both the power-spectrum and
  bi-spectrum can be calculated exactly analytically. The two-point
  function is characterised by different power on large and small
  scales, and a burst of small amplitude superimposed oscillations in
  between.  Motivated by string-theory models with stuck branes, we
  extend this analysis to Dirac Born Infeld (DBI) inflation, for which
  generalised slow-roll is violated at the discontinuity and a rapid
  variation in the speed of sound $\cs$ occurs. In an attempt to
  characterise the effect of non-linear kinetic terms on the
  oscillatory features of the primordial power-spectrum, we show that
  the resulting power spectrum has a shape and features which differ
  significantly from those of the standard Starobinsky model. In
  particular, the power-spectrum now takes very similar scale
  invariant values on large and small scales, while on intermediate
  scales it is characterised by much larger amplitude and higher
  frequency superimposed oscillations. We also show that calculating
  non-Gaussianities in this model is a complicated but interesting
  task since all terms in the cubic action now
  contribute. Investigating whether the superimposed oscillations
  could fit to the Planck Cosmic Microwave Background (CMB) data (for
  instance by explaining the large scale Planck anomalies) with, at
  the same time, small non-Gaussianities remains an intriguing and
  open possibility.
\end{abstract}

\maketitle

\section{Introduction}
\label{sec:intro}

It is by now a near certainty that the universe underwent a period of
accelerated expansion --- inflation --- early in its history. Indeed,
the spectacular Planck data \cite{Ade:2013rta,Ade:2013ydc} is entirely
compatible with Standard Single Field Inflation (SSFI) in the
slow-roll regime, and with a canonical kinetic
term~\cite{Martin:2013tda,Martin:2013nzq,Martin:2013gra}. However, the
nature of the inflaton field still remains a mystery, and on the
theoretical side much work has been carried out in the last years with
the aim of trying to embed the inflationary scenario in a high-energy
theory such as string theory. This paper fits into such a ``top-down''
approach.  In particular our focus is on string motivated models (see
below) in which, because generalised slow-roll is violated for a few
e-folds, oscillations are generated in the power spectrum. Our aim is
to determine precisely the observational predictions of this class of
models. In the process we will characterise how, for a given inflaton
potential, the properties of the superimposed oscillations in ${\cal
  P}_\zeta(k)$ --- amplitude and frequency for example --- depend on
non-standard kinetic terms in the Lagrangian for the inflaton.

The model we study is closely related to the so-called Starobinsky
model \cite{Starobinsky:1992ts,Starobinsky:1998mj} for which
the potential $V(\phi)$ is linear with a sharp change of slope at a
certain $\phi_0$:
\begin{equation} 
V(\phi) = \left\{\begin{array}{ll} \displaystyle
    V_0 + A_{+}\, \left(\phi-\phi_0\right)\ & {\rm for}\ \phi>\phi_0,\\
    \displaystyle V_0 + A_{-}\, \left(\phi-\phi_0\right)\ & {\rm for}\
    \phi<\phi_0.
\end{array}\right.
\label{Vstaro} 
\end{equation} 
In the following we will take $A_+>A_-$. The change in slope causes a
short, of order one in $e$-folds, period of fast roll, and remarkably
(in SSFI with standard kinetic terms) both the power spectrum and
bispectrum can be determined exactly analytically (in all range of
parameter space). To our knowledge, this is the only model in SSFI for
which any exact statements can be made.  One finds
\cite{Starobinsky:1992ts} that there is a sharp rise in the
power-spectrum ${\cal P}_\zeta(k)$ on scales $k\sim k_0$ (where $k_0$
is the mode that left the Hubble radius at $\phi=\phi_0$), with
\begin{equation} 
\lim_{k/k_0 \rightarrow 0} {\cal P}_\zeta^{_{\rm CS}}(k) =
\left( \frac{H_0}{2\pi} \right)^2 \left(\frac{3 H_0^2}{A_+}
\right)^2\, , 
\quad
\lim_{k/k_0 \rightarrow \infty} {\cal P}_\zeta^{_{\rm CS}}(k) = \left(
  \frac{H_0}{2\pi} \right)^2 \left(\frac{3 H_0^2}{A_-} \right)^2\, ,
\end{equation} 
where $H_0$ is the Hubble scale at the time when $\phi=\phi_0$, and CS
denotes the `canonical' Starobinsky model.  Thus the increase in power
is proportional to $A_-^{-2} - A_+^{-2}$, and it is followed by small
oscillations for $k\gsim k_0$ whose amplitude are rapidly damped out.
The different contributions to the bi-spectrum can also be calculated
analytically \cite{Martin:2011sn,Arroja:2012ae} in terms of $A_\pm$
and the third parameter of the model $H_0$, and some ranges of
parameter space are ruled out by recent constraints on $\fnl$ from
Cosmic Microwave Background (CMB) data
\cite{Ade:2013ydc,Hazra:2012yn}.

Here, motivated by string-theory, we focus on Dirac Born Infeld (DBI)
brane inflation (see \eg
\cite{Alishahiha:2004eh,Silverstein:2003hf,Lorenz:2007ze,Lorenz:2008je,Martin:2013uma,Langlois:2008qf})
with action
\begin{equation} 
S = -\int {\rm d}^4x \sqrt{-g}\left[ \Mp^2R + T(\phi)
  \sqrt{1+\frac{1}{T(\phi)}\,g^{\mu \nu}\partial _{\mu}\phi
  \partial _{\nu}\phi} 
  +V(\phi)-T(\phi) \right],
\label{eq:action}
\end{equation} 
where $R$ is the Ricci scalar and $\Mp$ the reduced Planck mass. As
usual $T(\phi)$, which we call the `brane tension' from the physical
origin of action (\ref{eq:action}), is determined by the warp factor
of the 10 dimensional metric in which the brane moves. Provided it is
continuous, its precise form is essentially unimportant for this
analysis, though to be concrete we take
\begin{equation} 
T(\phi) = \frac{\phi^4}{\lambda}
\label{Tdef}
\end{equation} 
which is typical for anti de Sitter warp factors.  The potential
$V(\phi)$ has contributions coming from the interaction between the
brane and the background, as well as any other branes which may be
present in the geometry. Its exact expression is not known and here,
without attempting to concretely realise the embedding of this model
within string-theory, we suppose it is given by (\ref{Vstaro}).  Such
a sudden jump at $\phi_0$ can be thought of as mimicking the presence
of a trapped brane stuck at a fixed point of an orbifold symmetry
\cite{Kofman:2004yc}, and potentials of this kind have been studied in
the DBI-literature before \eg \cite{Brax:2011si}. In particular, if
one considers the excitation of particles living on the trapped brane,
then these can backreact on the inflaton dynamics \cite{Brax:2009hd},
though the effect has been shown to be small.  Note that in the limit
in which the sound speed $\cs \rightarrow 1$, our model reduces to the
standard Starobinsky-model discussed above. Our model may not be
completely realistic from a string theory point of view but its
crucial advantage is that it allows us to derive explicit analytical
results, which can be compared to known results for the standard
Starobinsky model. We believe that the scenario studied here
represents the best compromise between cases in which the string model
building problem can be properly addressed (often at the expense of
solving the equations numerically), and over simplified situations in
which analytical results can be easily derived.

Just as in the standard Starobinsky model, in the DBI case,
generalised slow-roll is broken for order one efolds around
$\phi_0$. Thus $\cs$ also changes rapidly, and we expect oscillations
to be generated in the power-spectrum. How does the shape of ${\cal
  P}_\zeta(k)$ depend on the non-standard kinetic terms? Are the
amplitude and wavelength of the oscillations sensitive to the
non-linear structure of the Lagrangian?  In this paper not only do we
determine ${\cal P}_\zeta(k)$ numerically for all $\cs$, but we also
show that the model is essentially completely soluble analytically in
the $\cs \ll 1$ limit, in terms of the parameters of the potential
$A_\pm$. Due to the non-linear kinetic terms in the action, ${\cal
  P}_\zeta(k)$ differs significantly from that of the canonical SSFI
Starobinsky model discussed above. Indeed, the action given in
(\ref{eq:action}) now contains a second dimensionful potential
$T(\phi)$ and it is this, rather than $A_\pm$, which determines the
power-spectrum on small and large scales:
\begin{equation} 
\lim_{k/k_0 \rightarrow 0} {\cal P}_\zeta(k) = \left(
  \frac{H_0}{2\pi} \right)^2 \left(\frac{H_0^2}{T_0} \right)
= \lim_{k/k_0 \rightarrow \infty} {\cal P}_\zeta(k) 
\end{equation} 
where $T_0=T(\phi_0)$. As opposed to the canonical Starobinsky model
in which there is a sharp rise in power across $k_0$ if $A_- \ll A_+$,
in the DBI-Starobinsky model there is {\it no} rise in power for any
$A_\pm$. Rapid variations of $\cs$ occur when the field crosses
$\phi_0$, and these give rise to large amplitude, high frequency,
superimposed oscillations in ${\cal P}_\zeta(k)$ which we will discuss
in section \ref{sec:powerspectrum}. An interesting question is whether
or not these oscillations could fit the Planck data, for instance
explaining the large scale Planck anomalies, while also remaining
compatible with constraints on non-Gaussianities. Indeed, as we
discuss in the conclusions, an interesting new feature of this model
is that we expect non-Gaussianities to be sourced predominantly from
two (or more) coupled vertices, leading to a complicated structure.
While it might be expected that the constraints are strong at least
for $\cs \ll 1$, for larger $\cs$ the situation regarding
non-Gaussianities is much less clear while the large amplitude
superimposed oscillations remain in the power-spectrum.
  
In the context of SSFI, the development of models leading to
oscillations was motivated by observed features in the power-spectrum
of CMB temperature
fluctuations~\cite{Martin:2006rs,Martin:2003sg,Martin:2004iv,
  Martin:2004yi,Komatsu:2010fb}. Indeed, they may find their origin in
initial conditions, arising for instance from non-Bunch-Davies initial
conditions (see \eg
\cite{Martin:2000xs,Martin:2003sg,Brandenberger:2012aj}), or from
deviations from slow-roll in SSFI.  Amongst the models studied in the
literature are, for instance, potentials whose derivative is
discontinuous \cite{Starobinsky:1992ts,Martin:2011sn}, as well as
potentials which contain a step (\eg
~\cite{Starobinsky:1998mj,Chen:2008wn}), or a sinusoidal modulation
(e.g.~\cite{Flauger:2009ab,Chen:2010bka,Leblond:2010yq}).  Though
these features can fit data better than a nearly scale invariant
power-spectrum (see for example~\cite{Aich:2011qv,Ade:2013rta}), this
is at the expense of including extra parameters into the potential
meaning that the statistical significance of the features is not so
obvious~\cite{Martin:2010hh}.  Other than solving for the
power-spectrum numerically
\cite{Ringeval:2007am,Hazra:2010ve,Jain:2008dw}, some semi-analytical
methods have been developed \cite{Choe:2004zg,Dvorkin:2010dn} though
these are generally valid only in certain limiting cases, for example
if the step is small or if the scalar field and metric perturbations
decouple \cite{Leblond:2010yq}.

In models with non-standard kinetic terms, the consequences of rapid
variations in $\cs$ in $k$-inflation have been investigated (both for
the power-spectrum and bispectrum) using the effective field theory
formalism in \cite{Park:2012rh}. The ``generalised slow-roll
approximation" has been extended to $k$-inflation \cite{Hu:2011vr},
and applied to DBI-inflation in \cite{Miranda:2012rm}, though there
the authors considered a step-like feature in $T(\phi)$. Notice that
Ref.~\cite{Bean:2008na} also carries out a detailed analysis of the
signatures of step-like feature in both $T(\phi)$ and $V(\phi)$.

The structure of this paper is the following. In section
\ref{sec:background} we first discuss the background evolution of the
system and define the generalised slow-roll parameters (subsection
\ref{subsec:exact}). At this point, our discussion is for a general
potential $V(\phi)$.  In subsection \ref{subsec:application} we focus
on the Starobinsky model itself, with potentials $V(\phi)$ and
$T(\phi)$ given in Eqs.~(\ref{Vstaro}) and (\ref{Tdef}) respectively.
From the form of $V(\phi)$ we are able to derive exact results
regarding the behaviour of the system at the transition $\phi=\phi_0$,
and these enable us to determine the evolution of all the slow-roll
parameters analytically, even when slow-roll is violated. Our results,
valid in the DBI limit $\cs \ll 1$, are shown to match perfectly with
a full numerical solution of the background equations.  In section
\ref{sec:powerspectrum}, using these exact results, we show that the
calculation the power-spectrum ${\cal P}_\zeta(k)$ reduces to solving
single differential equation with particular time-dependent
coefficients that we specify.  The resulting power-spectrum is shown
to be in perfect agreement with a full numerical determination of the
same quantity.  We also determine analytically the dependence of
${\cal P}_\zeta(k)$ on $A_\pm$, on large and small scales. Finally we
summarise our main results in the section \ref{sec:conclusions}, where
we discuss in detail the different shape of the canonical and DBI
Starobinsky power-spectra. We also present a few considerations on the
calculation of non-Gaussianities in this model and mention interesting
directions for future work.

\section{Background equations and slowly varying parameters}
\label{sec:background}

\subsection{Exact Equations}
\label{subsec:exact}

For arbitrary potentials $V(\phi)$ and $T(\phi)$, and working in a
spatially flat Friedmann-Lema\^{\i}tre-Robertson-Walker (FLRW)
background geometry with metric $ \dd s^2 = -\dd t^2 + a^2(t)\dd {\bm
  x}^2 $, the Friedmann and scalar field equations of motion following
from Eq.~(\ref{eq:action}) are given, respectively, by
\begin{eqnarray} 
H^2 &=& \frac{1}{3\Mp^2} \left[(\gamma-1)T + V \right],
\label{FriedstDBI}
\\
\ddot{\phi} &+& \frac{3H}{\gamma^2}\dot{\phi} + 
\frac{3\gamma - \gamma^3 - 2}{2\gamma^3} T_\phi +
\frac{1}{\gamma^3} V_\phi = 0 \, .
\label{KGDBI}
\end{eqnarray} 
Here a dot/subscript $\phi$ denotes derivative with respect to cosmic
time $t$/field $\phi$ respectively, and $H=\dot{a}/a$ is the Hubble
parameter. The Lorentz factor $\gamma$ is related to the square-root
in (\ref{eq:action}) and is inversely proportional to the sound-speed
$\cs$
\begin{equation}
\gamma(\dot{\phi},\phi) \equiv \frac{1}{\cs } =
\frac{1}{\sqrt{1-\dot\phi^2/T(\phi)}} \, .
\label{gamma}
\end{equation} 
Later on we will mainly consider $\cs \ll 1$, which is the opposite
limit to the standard inflationary case, obtained when $\cs
\rightarrow 1$ (or $\gamma \rightarrow 1$). For the moment, however,
we leave $\gamma$ arbitrary and all the expressions in this section
are exact.

\par

In general, due to their complexity, Eqs.~(\ref{FriedstDBI})
and~(\ref{KGDBI}) cannot be integrated exactly unless numerical
methods are used. However, it is also interesting to have analytical
approximations and for this reason, we now define the horizon-flow
parameters. While in standard inflation they are defined as the
successive derivatives of the Hubble parameter \cite{Schwarz:2001vv,
  Leach:2002ar, Schwarz:2004tz}, in DBI-inflation a second hierarchy
of parameters must be introduced in order to describe the evolution of
the sound speed (or equivalently $\gamma$). This hierarchy is defined
as the successive derivatives of the Lorentz factor with respect to
the number of e-folds $N=\ln(a/a_\uin)$ (where $a_\uin$ is the initial
value of the scale factor), see e.g.~\cite{Chen:2006nt}. Thus the
slow-roll (or slowly-varying) parameters of DBI-inflation are given by
\begin{eqnarray} 
\epsilon_{n+1} &=& \frac{\dd \ln |\epsilon_n|}{\dd N},  \qquad 
\epsilon_0 \equiv \frac{H_\uin}{H}\, ,
\label{epsdef}
\\
\delta_{n+1} &=& \frac{\dd \ln |\delta_n|}{\dd N}, \qquad 
\delta_0 \equiv \frac{\cs{}_\uin}{\cs} = \frac{\gamma}{\gamma_\uin}.
\label{deldef}
\end{eqnarray}
The slow-roll approximation will consist in taking $|\epsilon_i | \ll
1$ and $|\delta_i |\ll 1$ and in sections \ref{subsec:application} we
will see that this greatly simplifies the equations describing the
evolution of the system.

For the moment, however, we make no approximation. In order to write
down the exact expressions for the first few slow-roll parameters, it
is useful to note from (\ref{FriedstDBI}) and (\ref{KGDBI}) that the
time derivative of the Hubble parameter is given by
\begin{equation}
\dot{H}=-\gamma \frac{\dot{\phi}^2}{2\Mp^2}, 
\end{equation} 
so that $H_\phi^2 =T(\phi)(\gamma^2 - 1)/(4\Mp^4)$. From here we can
extract a very useful expression for $\gamma$, namely
\begin{equation}
\gamma = \sqrt{1 + 4 \Mp^4 \frac{H_\phi^2}{T}}
\label{gammaex}
\end{equation}
which will be used extensively below, and in terms of which
derivatives of $N$ can easily be calculated
\begin{equation}
  \frac{\dd N}{\dd \phi}
  = -\frac{\gamma}{2\Mp^2}\frac{H}{ H_\phi}
  =-\frac{1}{2\Mp^2}\frac{H}{H_\phi}\sqrt{1+4 \Mp^4 \frac{H_\phi^2}{T}} \, .
\label{dNdphi}
\end{equation}
The exact expressions for the first flow parameters are then
\begin{eqnarray}
\epsilon_1 &=& \frac{2\Mp^2}{\gamma}\left( \frac{ H_\phi}{H} \right)^2 
= \frac{T(\gamma^2-1)}{2\Mp^2\gamma H^2} \, ,
\label{eps1}
\\
\epsilon_2 &=& -\frac{4 \Mp^2H_{\phi \phi}}{H\gamma} + 2 \epsilon_1 - \delta_1 \, ,
\label{eps2}
\\
\delta_1 &=& -\frac{2\Mp^2}{\gamma^2}
\frac{ H_\phi}{H}\frac{{\rm d}\gamma}{{\rm d}\phi} 
= -\frac{2\Mp^2}{\gamma^2}\frac{ H_\phi}{HT}
\left[ 6\Mp^2HH_\phi - V_\phi -\left(\gamma-1\right)T_\phi \right] \, ,
\label{del1}
\\
\delta_1 \delta_2 &=& \frac{3}{2} \left( \frac{\gamma^2-1}{\gamma^2} \right) 
\left(4\epsilon_1 - \epsilon_2-\delta_1 \right)
-\frac{3}{2}\epsilon_2\delta_1 +4\epsilon_1\delta_1
+ \frac{(\gamma^2+7)}{(\gamma^2-1)}\frac{\delta_1^2}{2}  
- \frac{(\gamma^2-1)}{\gamma^3 H^2} 
\left[V_{\phi \phi} + \left(\gamma-1\right)T_{\phi \phi} \right]\, ,
 \label{bloodymess}
\\
\epsilon_2\epsilon_3 +\delta_1 \delta_2 &=& -3(\epsilon_2 - 4\epsilon_1 + \delta_1) 
+ 2 \epsilon_1\epsilon_2 +  2\delta_1^2\left( \frac{\gamma^2}{\gamma^2-1}\right) 
- \frac{1}{2} (\epsilon_2 - 2\epsilon_1 + \delta_1) 
(\epsilon_2 - 4\epsilon_1 + 3\delta_1)
\nonumber \\ & &
- \frac{1}{\gamma H^2}\left[2V_{\phi \phi} 
+ \frac{T_{\phi \phi}}{\gamma}(\gamma-1)^2 \right]. 
 \label{bloodymessbis}
\end{eqnarray}
One can also proceed the other way round, and express some important
background quantities in terms of the slow-roll parameters. For
instance, on rewriting $T$ in terms of $\epsilon_1$ using
(\ref{eps1}), the Friedmann equation (\ref{FriedstDBI}) can be
re-expressed as
\begin{equation}
\label{eq:frieddbi}
H^2=\frac{V(\phi)}{3\Mp^2}\left[
1-\frac{2\gamma }{3\left(\gamma +1\right)}\epsilon_1\right]^{-1}.
\end{equation}
As we will see in the next sub-section, this equation turns out to be
crucial in order to understand the behaviour of the system. On
differentiating with respect to $\phi$ and using the definition of the
slow-roll parameters yields
\begin{equation}
\Mp\frac{V_\phi}{V} = +\sqrt{2\gamma\epsilon_1}
\left[1+ \frac{1}{3-2\gamma\epsilon_1/(\gamma+1)}
\frac{\gamma}{\gamma+1}
\left( \epsilon_2 + \frac{\delta_1}{\gamma+1} \right) \right] \, .
\label{exact}
\end{equation}
A further derivative would give $V_{\phi \phi}$, but the result is
somewhat tedious, so we only give it below to leading order in
slow-roll parameters. Finally, one can also derive a useful relation
for the derivative of the brane tension, namely
\begin{equation}
\Mp\frac{T_\phi}{T} = +\sqrt{\frac{\gamma}{2\epsilon_1}}
\left[2\epsilon_1 + \left( \frac{\gamma^2+1}{\gamma^2-1} \right) 
\delta_1 - \epsilon_2 \right] \, .
\label{Tder}
\end{equation}
As already mentioned, the above equations are all exact and valid for
any potential and any brane tension. In the next subsection, we focus
on the Starobinsky model itself.

\subsection{Application to the Starobinsky Model}
\label{subsec:application}

We now consider the DBI-Starobinsky model, with potentials $V(\phi)$
and $T(\phi)$ given in Eqs.~(\ref{Vstaro}) and (\ref{Tdef}). In this
case, Eqs.~(\ref{FriedstDBI}) and~(\ref{KGDBI}) still cannot be
integrated analytically. Nevertheless, as we now show, some exact
results about the behaviour of the system at the transition can be
established.

\subsubsection{Exact Results for Starobinsky Potential}
\label{subsubsec:exact}

Suppose that the field is rolling down the potential~(\ref{Vstaro}),
starting at a value $\phi_\uin$ with $\phi_\uin > \phi_0$. Since
$V_\phi$ is discontinuous at $\phi=\phi_0$ it follows from
Eq.~(\ref{KGDBI}) that both $\phi$ and $\dot{\phi}$ (and thus
$\gamma$) are continuous, but $\ddot{\phi}$ is discontinuous. Thus
$\dot{\gamma}$ is also discontinuous, and its jump can be read off
from the following exact equation [consequence of
Eqs.~(\ref{FriedstDBI}) and~(\ref{KGDBI})]
\begin{equation}
\dot{\gamma} = -\frac{\dot{\phi}}{T}\left[{3H}{\gamma}\dot{\phi}  
+ V_\phi +\left(\gamma-1\right)T_\phi \right] \, ,
\label{gammadot}
\end{equation}
which leads to
\begin{equation}
 \left[ \dot{\gamma}
\right]_\pm = -\frac{\dot{\phi}}{T} \left[ V_\phi \right]_\pm = \Delta
A \frac{\dot{\phi}}{T},
\label{jump1}
\end{equation}
where $ \Delta A \equiv A_- - A_+ < 0$. This can be rewritten in terms
of derivatives with respect to $N$ as
\begin{equation}
\left[ \frac{\dd \gamma}{\dd N} \right]_\pm = - \frac{\gamma^2 - 1}
{\Mp H^2 \gamma} \frac{\Delta A }{ \sqrt{2\epsilon_1 \gamma} }.
\label{jump2}
\end{equation}

We now study the behaviour of the slow-roll parameters at the
transition. From Eq.~(\ref{eps1}), it follows that $\epsilon_1$
remains continuous. However, since $\epsilon_2$ is defined in terms of
derivatives of $\epsilon_1$ which itself contains $\gamma$ [see
Eq.~(\ref{eps1})], the second horizon flow parameter $\epsilon_2$ is
discontinuous.  (This is also true in the the SSFI-Starobinsky model.)
Moreover, following from the definition in terms of $\gamma$, the
parameter $\delta_1$ is also discontinuous. Its jump across the
discontinuity is related to that of $\epsilon_2$ from
Eq.~(\ref{Tder}):
\begin{equation}
\label{jumpd1}
\left[\delta_1 \right]_\pm \left. 
\left( \frac{\gamma^2+1}{\gamma^2-1} \right)
\right|_{\phi_0}  =  \left[\epsilon_2 \right]_\pm .
\end{equation}
If initial conditions at the beginning of inflation are such that all
slow-roll parameters are small, it therefore follows that the
DBI-Starobinsky model is characterised by a continuous $\epsilon_1$
which remains small all the time (hence, inflation never comes to an
end), and by parameters $\epsilon_2$ and $\delta_1$ which are small
far from the transition, but jump and can be large at the
discontinuity. In this sense, the DBI-Starobinsky model is a direct
generalisation of the canonical Starobinsky model for which
$\epsilon_1\ll 1$ and $\epsilon_2$ jumps at the transition. The new
ingredient is, of course, the presence of the parameter $\delta_1$ and
we have just seen that this parameter has a jump comparable to that of
$\epsilon_2$, in particular when $\gamma \gg 1$, see
Eq.~(\ref{jumpd1}).

\begin{figure*}
\includegraphics[width=0.45\textwidth,height=.45\textwidth]{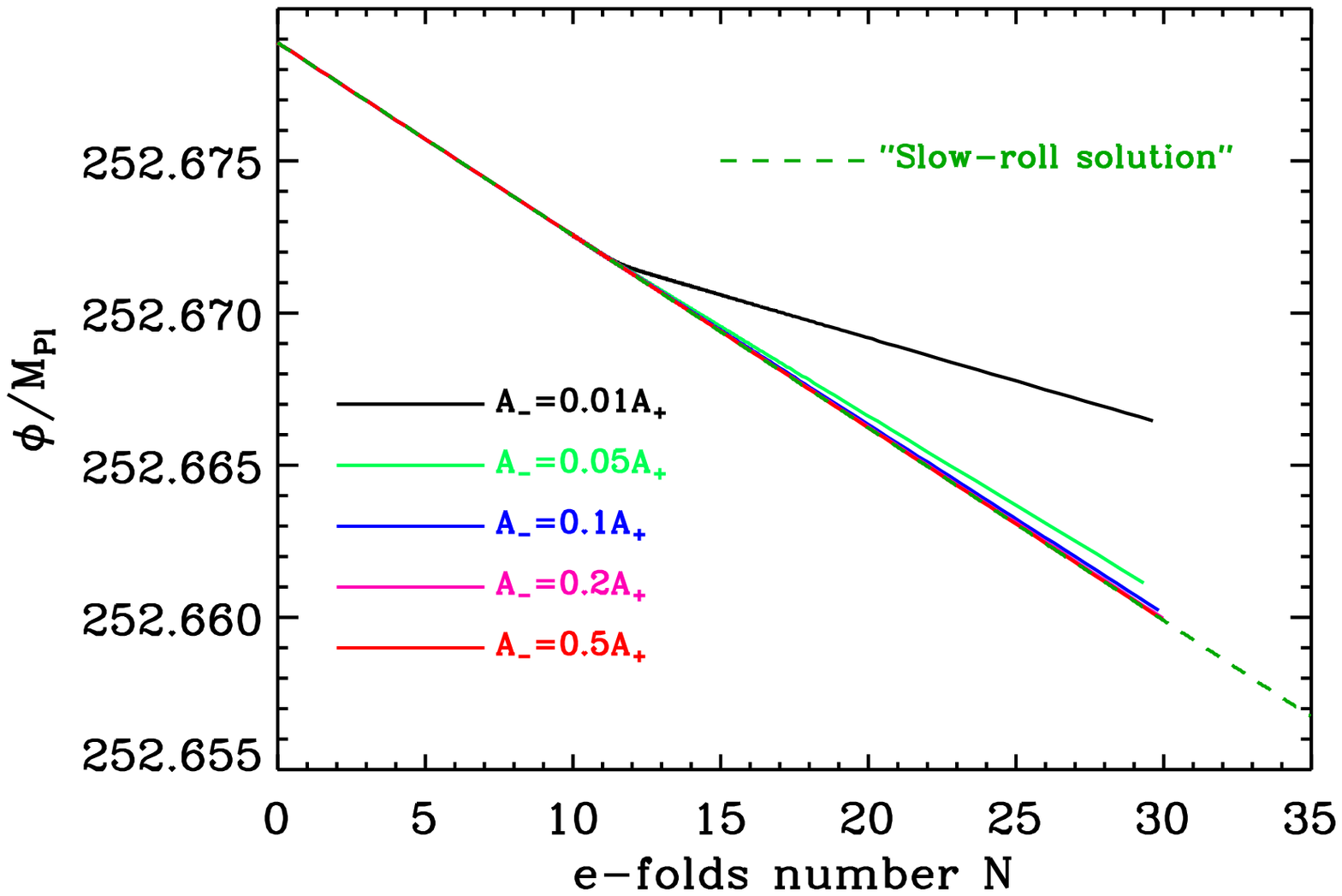}
\includegraphics[width=0.45\textwidth,height=.45\textwidth]{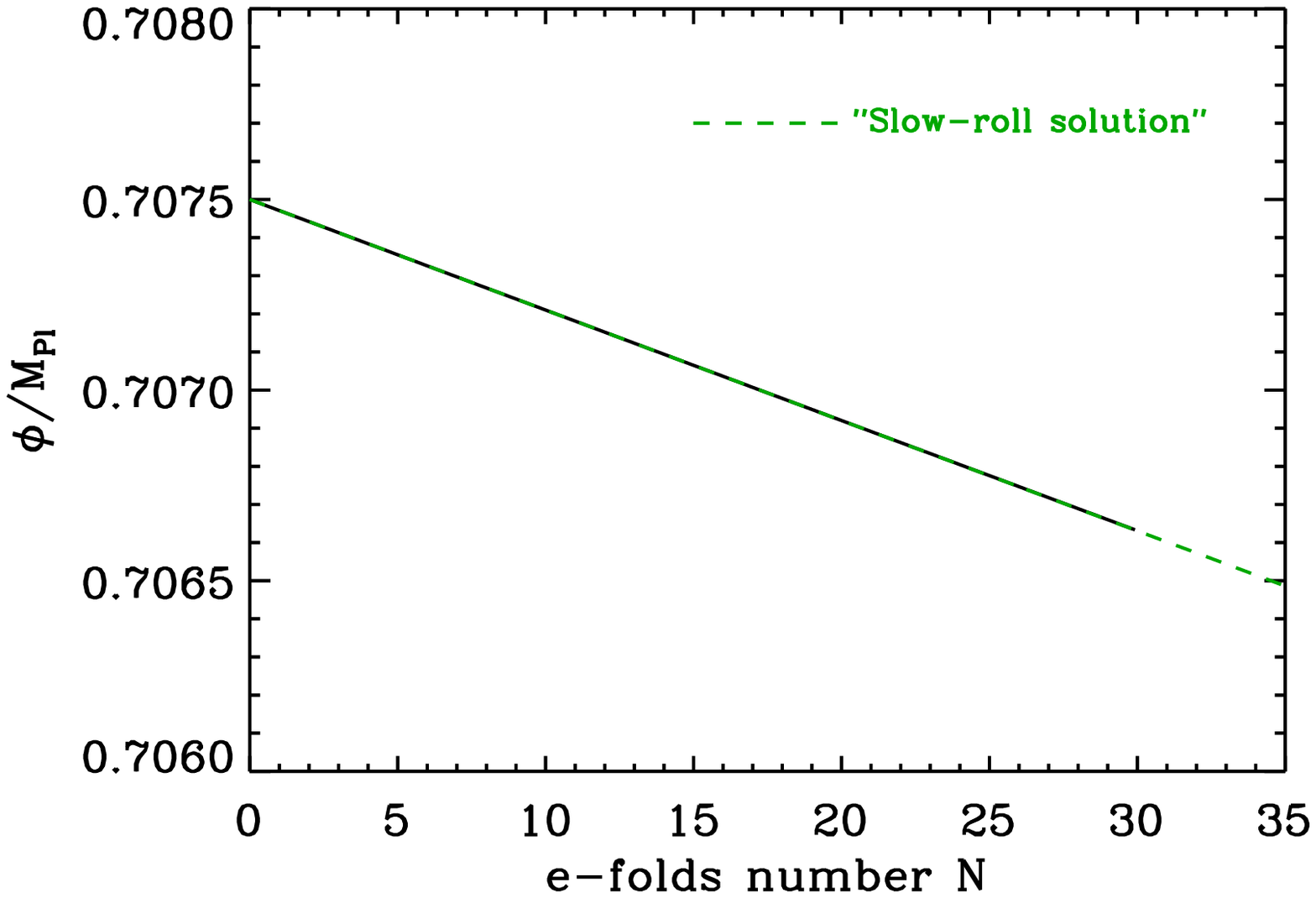}
\caption{{\bf Left panel}: Evolution of the field $\phi$ for the
  choice $\epsilon_{1\uin}\simeq 10^{-5}$, $\epsilon_{2\uin}\simeq
  2.5\times 10^{-5}$, $\delta _{1\uin}\simeq 1.5\times 10^{-5}$,
  $\gamma _{\uin}\simeq 50$, $\phi_\uin/\Mp\simeq 252.6788$,
  $H_\uin/\Mp \simeq 3.97\times 10^{-8}$ and $N_0=10$, where $N_0$ is
  the number of e-folds at which the field goes through the
  transition. This implies $V_0/\Mp^4 \simeq 4.71 \times 10^{-15}$,
  $\lambda \simeq 6.45\times 10^{30}$, $\phi_0/\Mp\simeq 252.6725$ and
  $A_+/\Mp^3\simeq 1.49 \times 10^{-16}$. In the following this set of
  parameters is named ``run one". The black solid curve corresponds to
  $A_-=0.01 A_+$, the solid green one to $A_-=0.05 A_+$, the solid
  blue one to $A_-=0.1 A_+$, the solid pink one to $A_-=0.2 A_+$ and,
  finally, the solid red one to $A_-=0.5 A_+$. The presence of the
  transition at $N_0=10$ is easily visible. We see that the greater
  the change in the slopes, the more the trajectory is modified after
  the transition. The dark green dashed line represents the slow-roll
  solution given by Eq.~(\ref{eq:fieldsr}). The slow-roll trajectory
  is not affected by the transition because, in the approximation used
  there, this trajectory no longer depends on the parameters
  $A_{\pm}$. If the initial velocity is not very large and the slope
  change important, then the actual trajectory can significantly
  deviates from the slow-roll, see the example of the black solid
  curve. Otherwise, the agreement is excellent.  {\bf Right panel}:
  evolution of the scalar field for another set of parameters, namely
  $\epsilon_{1\uin}\simeq 7.25\times 10^{-7}$, $\epsilon_{2\uin}\simeq
  -7.98\times 10^{-5}$, $\delta _{1\uin}\simeq 8.27\times 10^{-5}$,
  $\gamma _{\uin}\simeq 1723.33$, $\phi_\uin/\Mp\simeq 0.7075$,
  $H_\uin/\Mp\simeq 1.82\times 10^{-9}$, $N_0\simeq 17.24$. This
  implies that $V_0/\Mp^4 \simeq 9.98 \times 10^{-18}$, $\lambda
  \simeq 8.95\times 10^{25}$, $\phi_0/\Mp\simeq 0.7070$,
  $A_+/\Mp^3\simeq 4.98 \times 10^{-19}$ and $A_-=0.01 A_+$. In the
  following, we denote this set of parameters by ``run two".  The
  solid black line represents the exact trajectory while the dark
  green dashed line represents the slow-roll solution given by
  Eq.~(\ref{eq:fieldsr}). Despite the fact that $A_-=0.01 A_+$, as for
  the solid black line in the left panel, the agreement between the
  numerical and slow-roll solutions, is now very good. This is due to
  the fact that, for run two, the Lorentz factor $\gamma $ is
  larger. As a consequence, the field arrives at the transition with a
  higher velocity and, therefore, is less sensitive to the changes in
  the slopes.}
\label{fig:field}
\end{figure*}

\subsubsection{Integration of Equations of Motion: Numerics and
  Analytic Approximation}
\label{subsubsec:numerics}

Having understood the broad behaviour of the background quantities
(without using any approximation), we now aim to understand their
evolution in a more detailed fashion, at the quantitative level. As
already mentioned, since Eqs.~(\ref{FriedstDBI}) and~(\ref{KGDBI})
cannot be integrated exactly analytically, we have to rely either on
numerical calculations or on approximations. In the following, we use
both.

Let us start with the field $\phi$. In Fig.~\ref{fig:field}, we show
the exact evolution of the field (we have numerically integrated the
exact equations of motion) for two sets of parameters and initial
conditions. In order to understand this behaviour, first notice from
Eq.~(\ref{eq:frieddbi}) that in the slow-roll regime when
$\epsilon_1\ll 1$, and for all $\gamma$, the Friedmann
equation~(\ref{eq:frieddbi}) reduces to
\begin{equation}
H^2\simeq \frac{1}{3\Mp^2}V(\phi).
\label{FF}
\end{equation}
It is remarkable that, despite its intrinsic complexity in DBI
inflation, the Friedmann equation exactly reduces to its standard
counterpart when $\epsilon_1\ll 1$. 
Thus we also have
\begin{equation}
 2\Mp \frac{H_\phi}{H } \simeq \Mp \frac{V_\phi}{V} \simeq
+\sqrt{2\gamma\epsilon_1}  \, .
\label{VV}
\end{equation}
Now, the dynamics of $\phi(N)$ can be obtained from the exact identity
(\ref{dNdphi}) which, for a slow-roll trajectory [using
Eqs.~(\ref{FF}) and~(\ref{VV})], reduces to
\begin{equation}
\label{eq:intsrtrajec}
N(\phi)\simeq \pm \frac{1}{\Mp^2}\int _{\phi_\uin}^{\phi}{\rm d}\psi
\sqrt{\left(\frac{V}{V_\psi}\right)^2+\Mp^2\frac{V}{3T}}.
\end{equation}
Notice that this expression is in fact valid for any potential and any
brane tension provided the slow-roll approximation holds. It was
established for the first time in
Ref.~\cite{Lorenz:2008et}. 

\par

For the DBI-Starobsinky model with potential given in (\ref{Vstaro}),
Eq.~(\ref{eq:intsrtrajec}) yields
\begin{equation}
  N(\phi)=- \frac{1}{\Mp^2}\int _{\phi_\uin}^{\phi}{\rm d}\psi
  \frac{V_0+A_{\pm}(\psi-\phi_0)}{A_{\pm}}
  \sqrt{1+\Mp^2\frac{\lambda A_{\pm}^2}{3\psi^4
      \left[V_0+A_{\pm}(\psi-\phi_0)\right]}}
  \nn,
\end{equation}
where we have chosen the minus sign since for the Starobinsky
potential $N$ increases as the field rolls down the potential towards
smaller $\phi$. Notice that the origin of the square-root is the
$\gamma$ factor in (\ref{dNdphi}). The above expression is still too
complicated to allow an exact integration to determine the field
trajectory: further assumptions must be made, and the first we make is
to assume vacuum domination $V_0 \gg A_\pm (\psi - \phi_0)$ for all
$\psi$. Then the trajectory is given by
\begin{equation}
N(\phi)\simeq - \frac{1}{\Mp^2}\frac{V_0}{A_{\pm}}
\int _{\phi_\uin}^{\phi}{\rm d}\psi
  \sqrt{1+\frac{\lambda A_{\pm}^2}{3V_0\Mp^2}
\left(\frac{\Mp}{\psi}\right)^4}
\label{chr}
\end{equation}
which results in Elliptic functions. Since this is still not
especially illuminating, we make a second assumption, namely that we
work in the DBI regime $\gamma \gg 1$, or equivalently $\cs \ll 1$.
Referring to (\ref{gammaex}) this implies that we neglect the ``1" in
$\gamma$, which in Eq.~(\ref{chr}) translates into
\begin{equation}
  N(\phi)\simeq -  \int _{\phi_\uin}^{\phi}{\rm d} \psi \frac{H}{\sqrt{T}} =
  - \frac{1}{\Mp} \sqrt{\frac{V_0 \lambda}{3}} 
\int _{\phi_\uin}^{\phi}{\rm d} \psi\frac{1}{\psi^2} \, .
  \nn
 \end{equation}
The solution is
\begin{equation}
  \frac{1 }{\phi(N)} =\frac{1}{\phi_\uin} -  \Mp  \sqrt{\frac{3}{\lambda V_0}} N\, 
  =\frac{1}{\phi_0} -  \Mp \sqrt{\frac{3}{\lambda V_0}}  (N-N_0) \, ,
\label{eq:fieldsr}
\end{equation}
where $N_0$ denotes the number of e-folds when $\phi$ reaches the
transition at $\phi_0$. The shortcoming of this expression is that,
because of our successive approximations --- vacuum domination,
slow-roll $\epsilon_1 \ll 1$, and DBI-regime $\gamma \gg 1$ --- we
have lost the dependence on the coefficients $A_{\pm}$. The exact
evolution of the scalar field is compared to the slow-roll
trajectory~(\ref{eq:fieldsr}) in Fig.~\ref{fig:field}.

\begin{figure*}
\includegraphics[width=0.45\textwidth,height=.45\textwidth]{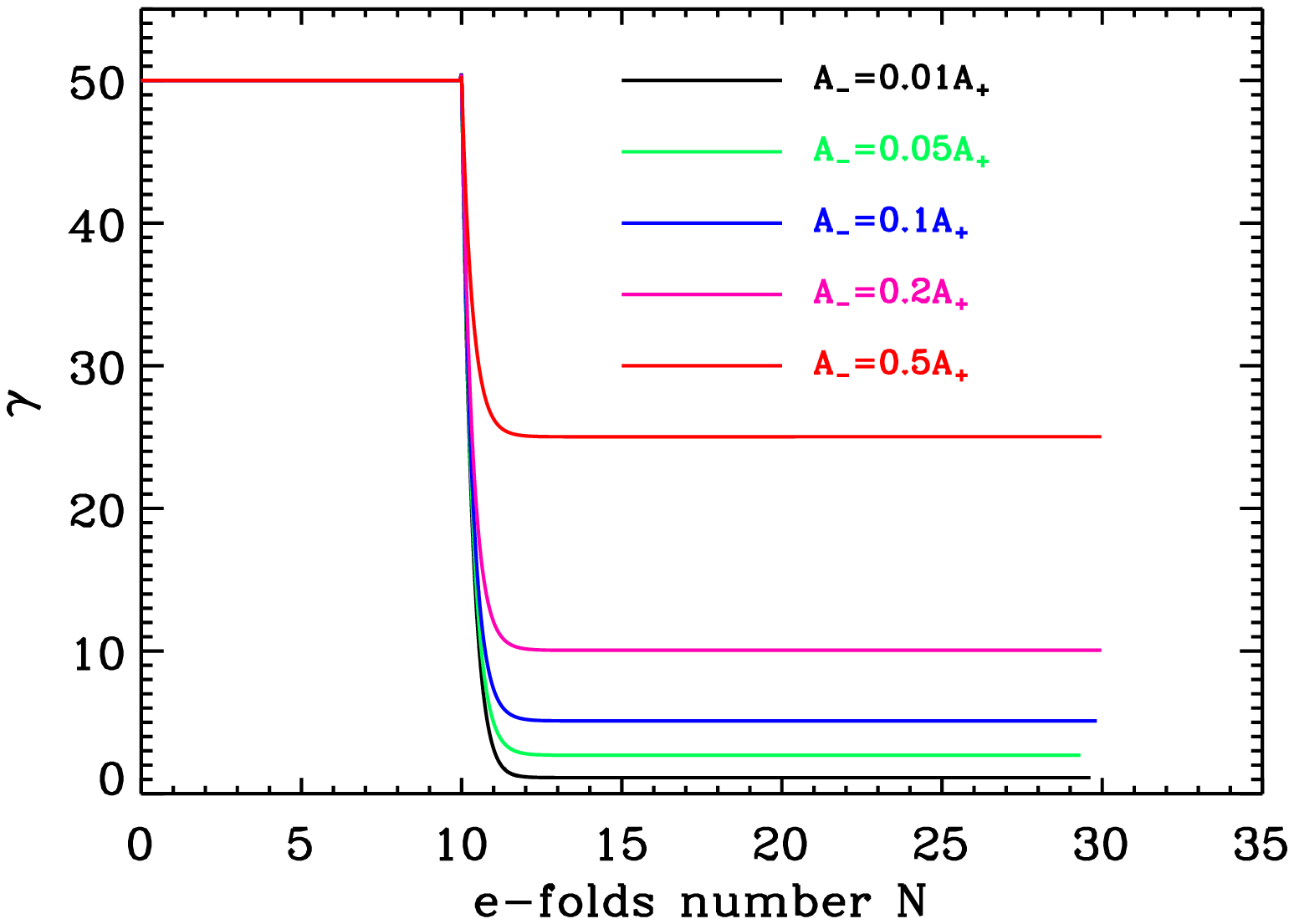}
\includegraphics[width=0.45\textwidth,height=.45\textwidth]{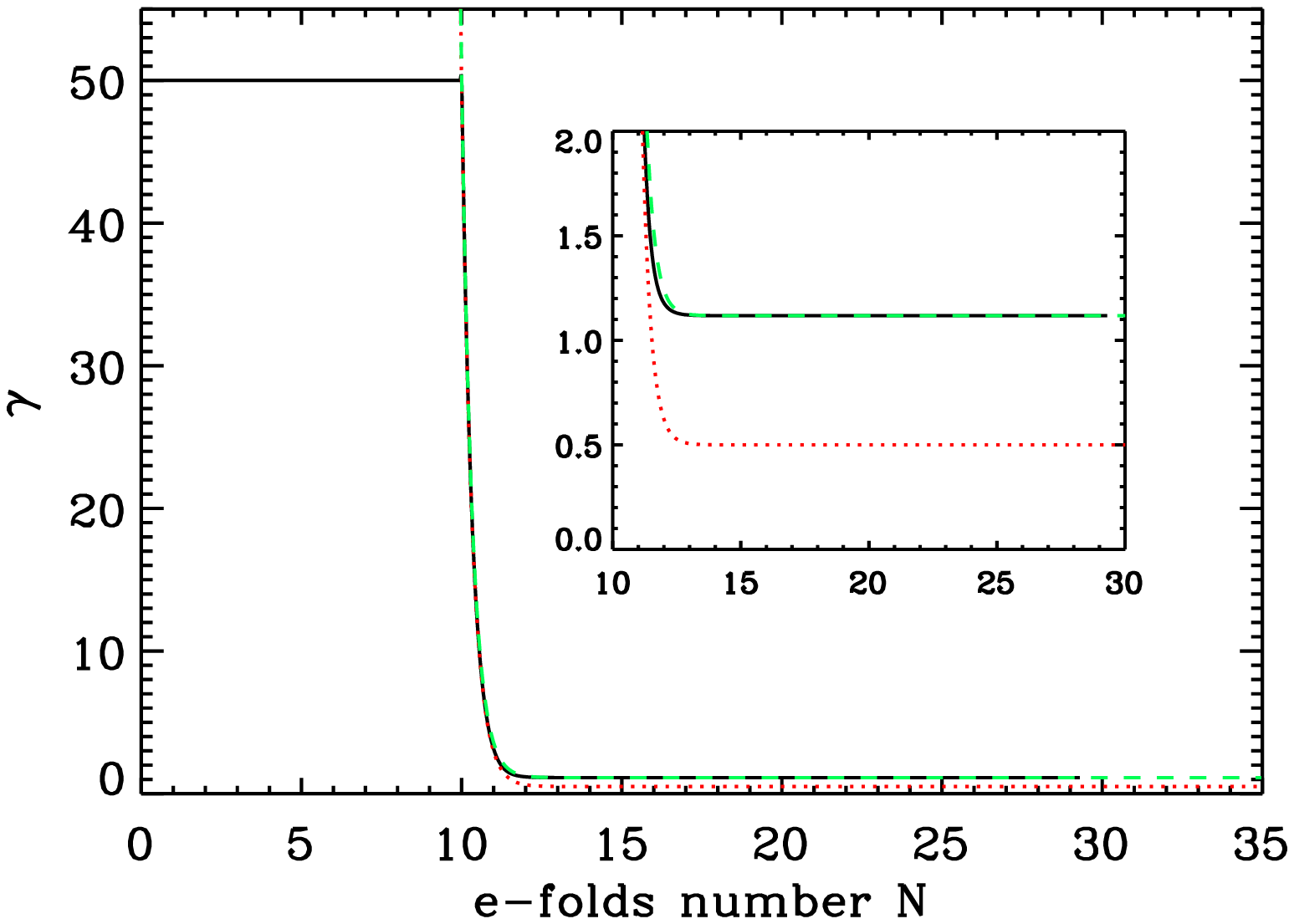}
\includegraphics[width=0.45\textwidth,height=.45\textwidth]{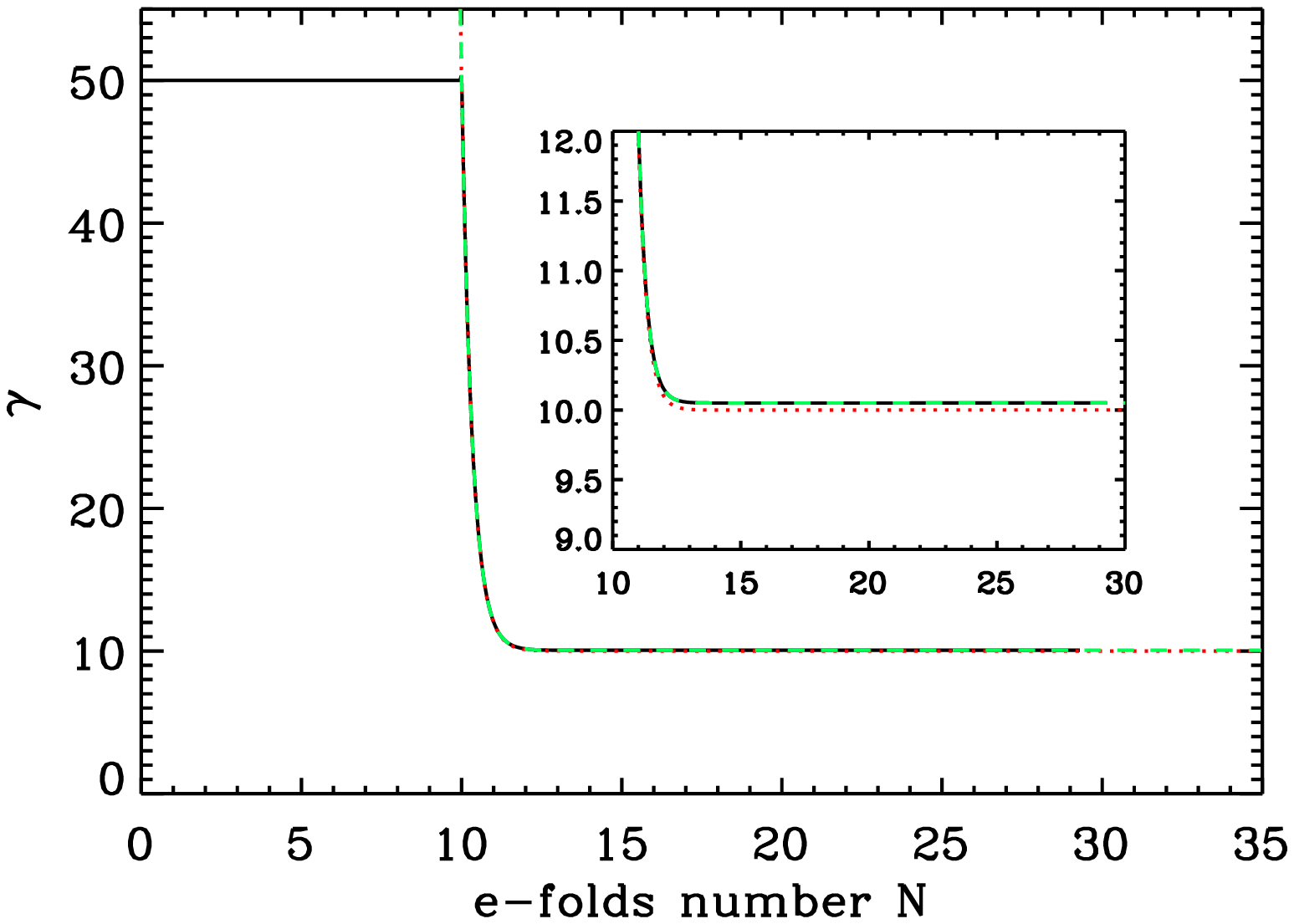}
\includegraphics[width=0.45\textwidth,height=.45\textwidth]{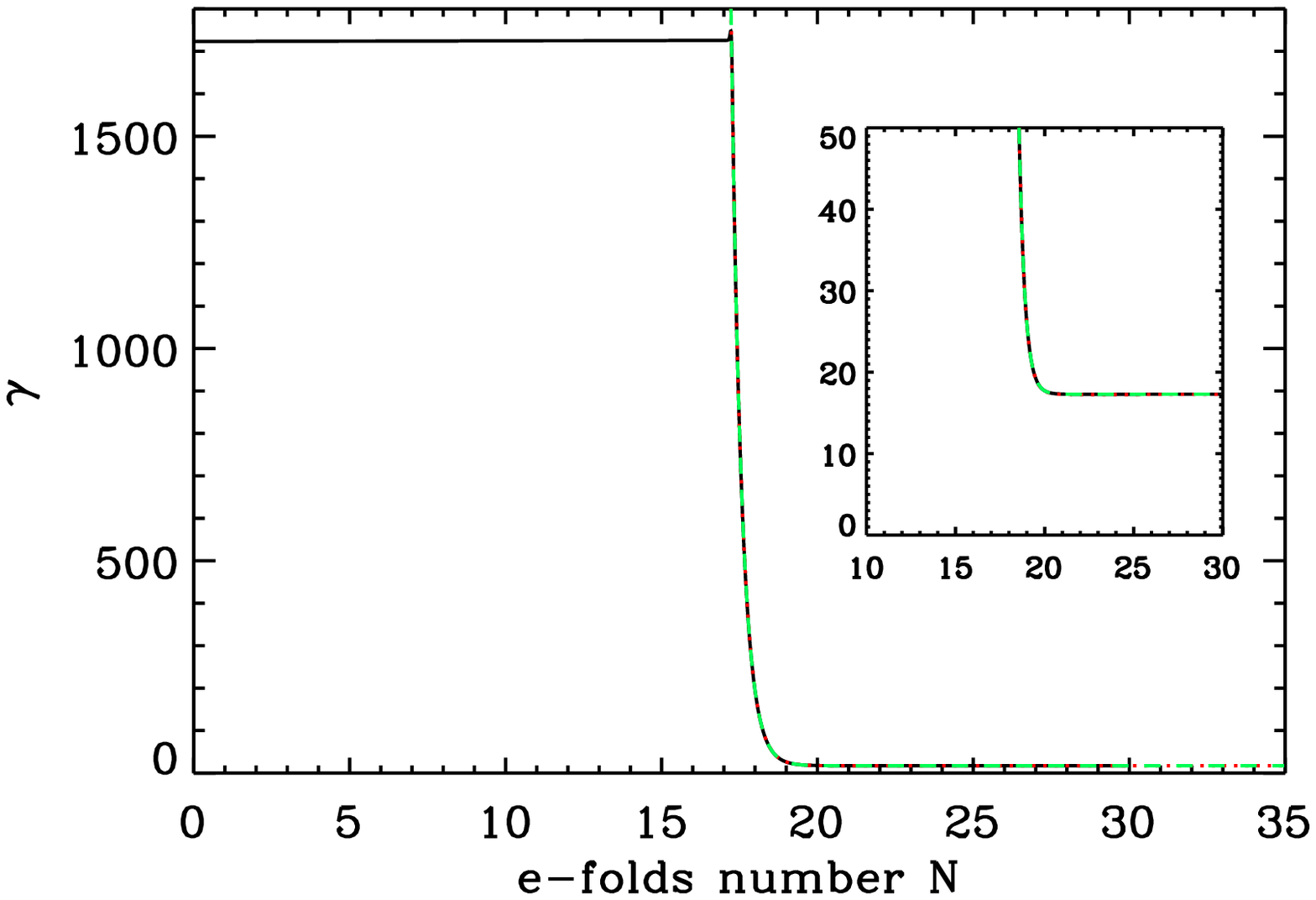}
\caption{{\bf Top left panel}: Evolution of the Lorentz factor $\gamma
  (N)$ for ``run one" (see the definition of ``run one" in the caption
  of Fig.~\ref{fig:field}) for $A_-=0.01 A_+$ (solid black line),
  $A_-=0.05 A_+$ (solid green line), $A_-=0.1 A_+$ (solid blue line),
  $A_-=0.2 A_+$ (solid pink line) and $A_-=0.5 A_+$ (solid red
  line). The transition at $N_0=10$ due to the discontinuity in the
  slope of the potential is clearly visible in this figure. {\bf Top
    right panel}: ``run one" for $A_-=0.01 A_+$ (same as the solid
  black line in top left panel). The dashed green curve corresponds to
  the approximate slow-roll evolution of the Lorentz factor given by
  Eq.~(\ref{eq:gammasr}) and valid after the transition. The dotted
  red curve corresponds to neglecting the term one inside the square
  in this expression and, therefore, to assuming that $\gamma _{_{\rm
      SR}}(N)=\sqrt{\lambda/(3V_0)}A_-\Mp/\phi^2(N)$. As shown in the
  inset, the dashed green line is an excellent fit while the dotted
  red line is not accurate enough. This is because the case $A_-=0.01
  A_+$ corresponds to a brutal change in the slope of the potential
  such that the field velocity strongly decreases after the
  transition. As a consequence, the Lorentz factor approaches one and
  the factor one in the square root in Eq.~(\ref{eq:gammasr}) can no
  longer be neglected. {\bf Bottom left panel}: same as top right
  panel but with $A_-=0.2 A_+$. As shown in the inset, this time, both
  the dashed green line and the dotted red line are good fits of the
  numerical solution. Clearly, this is because the change of slopes is
  less abrupt and, therefore, the field velocity decreases less at the
  transition. As a consequence, the Lorentz factor remains large
  compared to one and the factor one in the in the square root in
  Eq.~(\ref{eq:gammasr}) can now be safely neglected. {\bf Bottom
    right panel}: same as top right panel but for ``run two" (see the
  definition of ``run two" in the caption of
  Fig.~\ref{fig:field}). The dashed green line and dotted red line are
  excellent fit of the actual numerical solution (see the inset)
  despite the fact that the change in the slopes is abrupt, $A_-=0.01
  A_+$. The reason for this behaviour is of course that the initial
  value of the Lorentz factor is higher.}
\label{fig:gamma}
\end{figure*}

\begin{figure*}
\includegraphics[width=0.45\textwidth,height=.45\textwidth]{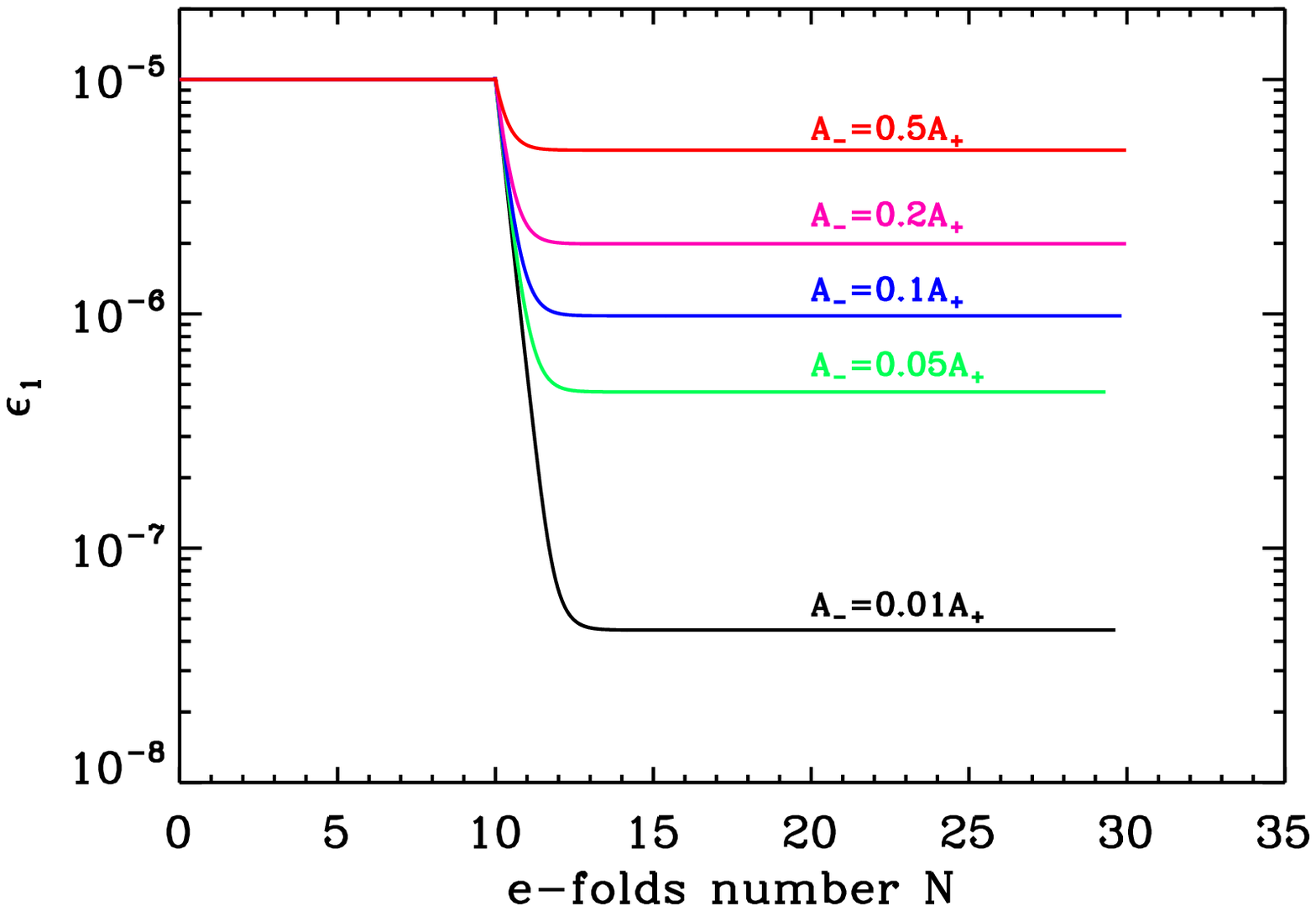}
\includegraphics[width=0.45\textwidth,height=.45\textwidth]{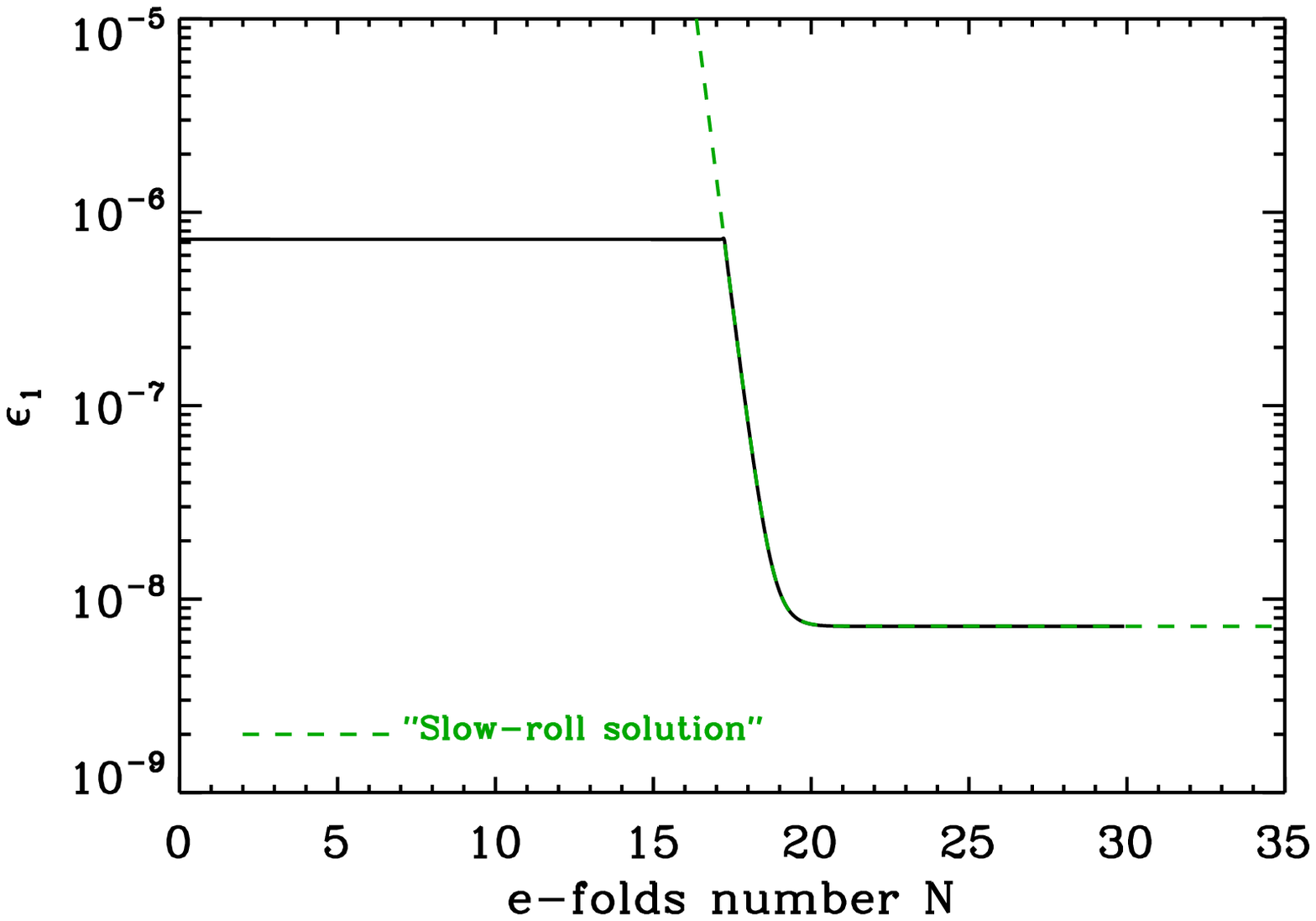}
\caption{{\bf Left panel}: Evolution of the first slow-roll parameter
  $\epsilon_1$ for ``run one" (see the definition of ``run one" in the
  caption of Fig.~\ref{fig:field}) for $A_-=0.01 A_+$ (solid black
  line), $A_-=0.05 A_+$ (solid green line), $A_-=0.1 A_+$ (solid blue
  line), $A_-=0.2 A_+$ (solid pink line) and $A_-=0.5 A_+$ (solid red
  line). {\bf Right panel}: Evolution of the first slow-roll parameter
  $\epsilon_1$ for ``run two'' (see the definition of "run two" in the
  caption of Fig.~\ref{fig:field}). The dark green dashed line
  corresponds to the ``slow-roll" solution given by
  Eq.~(\ref{eps1fit}) (and valid only after the transition). As can
  be observed from the figure, this is an excellent fit to the
  numerical solution.}
\label{fig:eps1}
\end{figure*}

We now study Lorentz factor $\gamma$ given in Eq.~(\ref{gammaex}) in
more detail. Its exact (numerically solved) evolution is shown in
Fig.~\ref{fig:gamma}. If the slow-roll approximation is satisfied
(that is to say far from the point where the derivative of the
potential is discontinuous), then
\begin{equation}
\label{eq:gammasr}
\gamma _{_{\rm SR}}(\phi)\simeq \sqrt{1+\frac{\lambda \Mp^2A_{\pm}^2}{3\phi^4
\left[V_0+A_{\pm}(\phi-\phi_0)\right]}}\simeq 
\sqrt{1+\frac{\lambda \Mp^2A_{\pm}^2}{3\phi^4
V_0}}\, ,
\end{equation}
where the last expression is valid in the vacuum dominated
regime. Furthermore, in the DBI regime $\gamma \gg 1$, the second term
in the square root must dominate so that
\begin{equation}
  \gamma _{_{\rm SR}}^{(\pm)}(N)\simeq  \sqrt{\frac{\lambda}{3V_0}}
  \frac{\Mp A_{\pm}}{\phi^2(N)},
  \label{gpma}
\end{equation}
where $\phi(N)$ is given in (\ref{eq:fieldsr}). Notice that the
Lorentz factor does depend on $A_\pm$.

During the transition, the above expression is clearly no longer valid
since slow-roll is violated. But we can determine the evolution of
$\gamma$ through the transition using Eq.~(\ref{jump2}) which, in the
$\gamma \gg 1$ limit, reduces to
\begin{equation} 
\left[ \frac{\dd \gamma}{\dd N}
\right]_\pm \simeq 3 \gamma_{_{\rm SR}}^{(+)}(N_0) \frac{\Delta A }{A_+} \, .
\label{jump3}
\end{equation}
Therefore, during the transition era, the Lorentz factor decreases
exponentially (recall that $\Delta A < 0$) and
\begin{equation}
\gamma^{(-)}(N) = \left[\gamma_{_{\rm SR}}^{(+)}(N_0) - \gamma_{_{\rm SR}}^{(-)}(N_0)
\right]{\rm e}^{-3(N-N_0)} + \gamma_{_{\rm SR}}^{(-)}(N)
\label{eq:gamev}
\end{equation}
since the continuity of $\gamma$ requires that $\gamma^{(-)}(N_0)
=\gamma_{_{\rm SR}}^{(+)}(N_0)$, while $\gamma(N\gg N_0)=\gamma_{_{\rm
    SR}}^{(-)}(N)$. Once again, this excellent fit for $\gamma$ is
shown in Fig.~\ref{fig:gamma}.

\begin{figure*}
\includegraphics[width=0.45\textwidth,height=.45\textwidth]{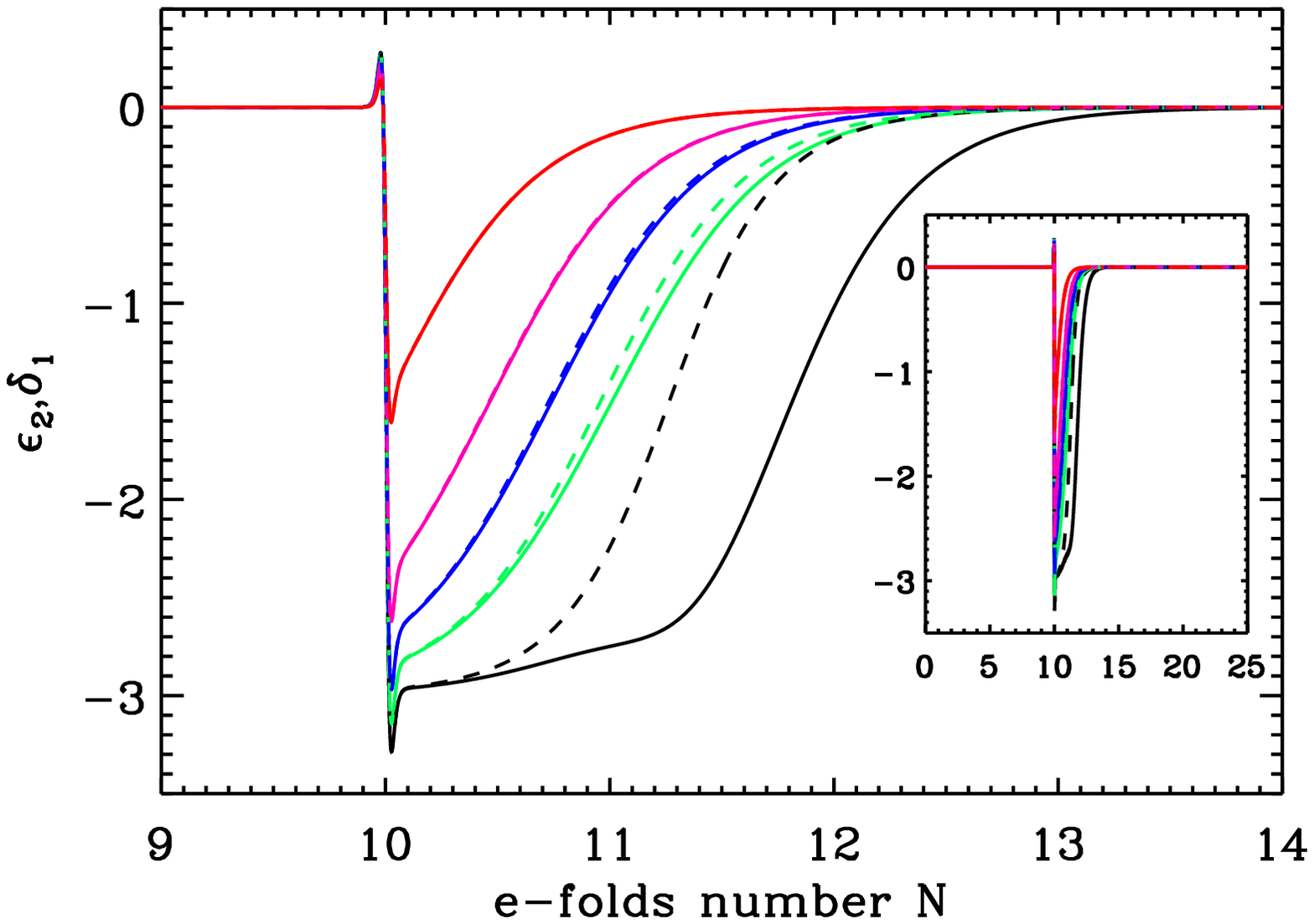}
\includegraphics[width=0.45\textwidth,height=.45\textwidth]{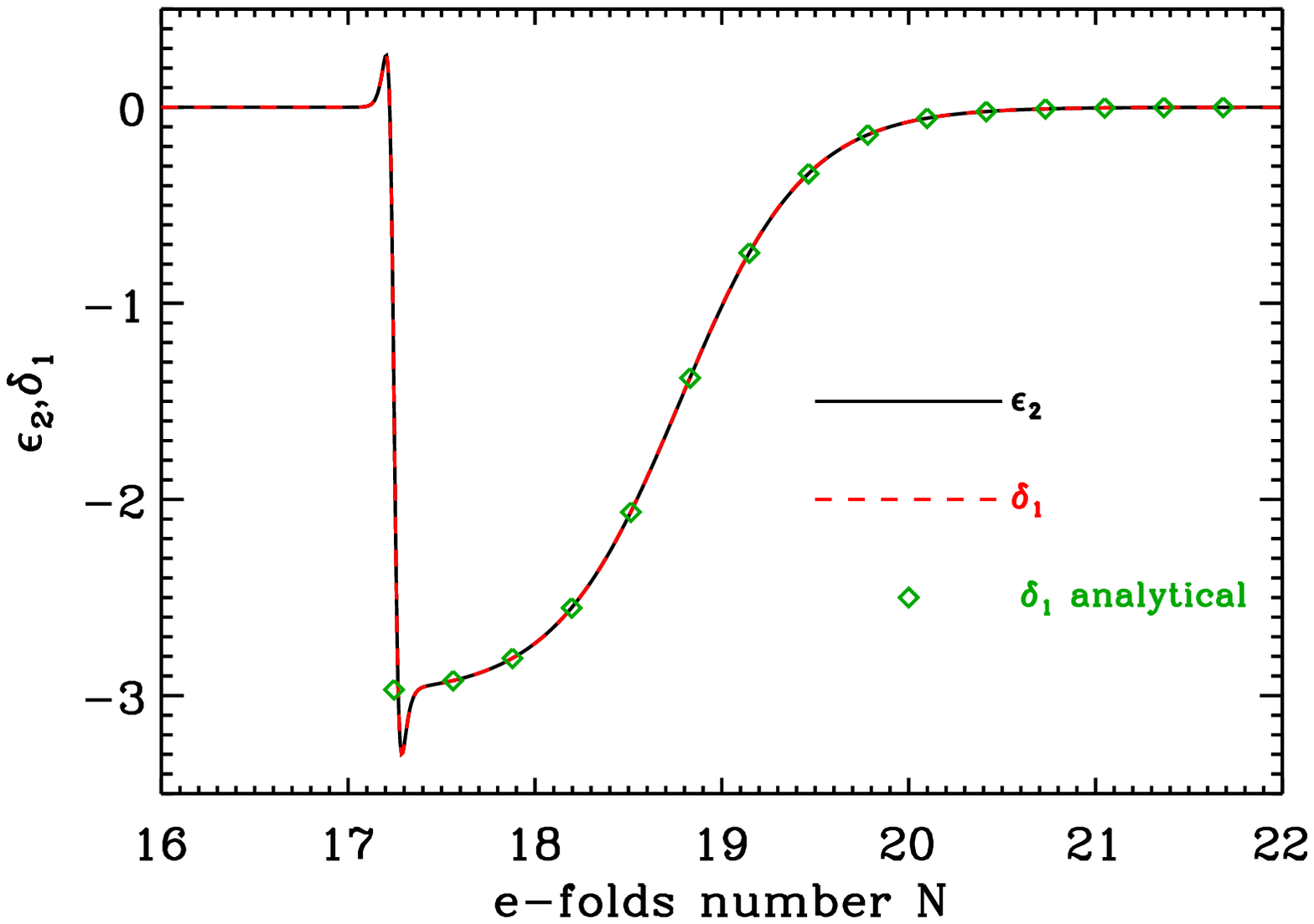}
\caption{{\bf Left panel}: evolution of the slow-roll parameters
  $\epsilon_2$ and $\delta _1$ for ``run one" (see the definition of
  ``run one" in the caption of Fig.~\ref{fig:field}) for $A_-=0.01
  A_+$ (solid black line for $\epsilon_2$ and dashed black line for
  $\delta _1$), $A_-=0.05 A_+$ (solid green line for $\epsilon_2$ and
  dashed green line for $\delta _1$), $A_-=0.1 A_+$ (solid blue line
  for $\epsilon_2$ and dashed blue line for $\delta _1$), $A_-=0.2
  A_+$ (solid pink line for $\epsilon_2$ and dashed pink line for
  $\delta _1$) and $A_-=0.5 A_+$ (solid red line for $\epsilon_2$ and
  dashed red line for $\delta _1$). Outside the region of the
  transition, around $N_0=10$, the evolution is featureless as shown
  in the inset. When the change in the slope is not too abrupt, one
  observes that $\epsilon_2\simeq \delta_1$. Of course the most
  important result shown in the figure is that $\epsilon_2$ and
  $\delta _1$ can be large during the transition. When the kinetic
  term is standard, slow-roll violation corresponds to a situation
  where $\epsilon_1\ll 1$ (\ie inflation never stops) and $\vert
  \epsilon_2\vert>1$. In the DBI case, this situation generalises to
  $\epsilon_1\ll 1$, $\vert \epsilon_2\vert >1$ and $\vert
  \delta_1\vert>1$. {\bf Right panel}: evolution of the slow-roll
  parameters $\epsilon_2$ and $\delta _1$ for ``run two" (see the
  definition of ``run two" in the caption of Fig.~\ref{fig:field}). In
  this case we have $\epsilon_2\simeq \delta_1$ with a very good
  approximation despite the fact that $A_-=0.01 A_+$. As before, this
  is due to the fact that the initial velocity of the field is much
  larger than in the left panel. The dark green dots correspond to the
  ``slow-roll" solution given by Eq.~(\ref{eq:delta1}) (and valid only
  after the transition). As can be noticed in the figure, this is an
  excellent fit to the numerical solution. The small peaks around
  $N_0$ are simply numerical artifacts whose origin stems from the
  fact that we have modelled the Heaviside function with a hyperbolic
  tangent. The amplitude of these peaks can be decreased at will by
  increasing the sharpness of the hyperbolic tangent.}
\label{fig:delta1}
\end{figure*}

We now turn to the behaviour of the slow-roll parameters. The quantity
$\epsilon_1$ is determined directly from Eq.~(\ref{eps1}). Away from
the transition, where the slow-roll approximation is valid, one
obtains
\begin{equation}
\epsilon_1{}_{_{\rm SR}}^{(\pm)}(N)\simeq \frac{\Mp A_{\pm}}{2V_0}
\sqrt{\frac{3}{\lambda V_0}}\phi^2(N),
\label{eq:eps1SR}
\end{equation}
while during the transition Eq.~(\ref{eps1}) yields
\begin{equation}
\epsilon_{1}^{(-)}(N) = \left[\epsilon_{1}{}_{_{\rm SR}}^{(+)}(N_0) 
-\epsilon_{1}{}_{_{\rm SR}}^{(-)}(N_0)\right]{\rm e}^{-3(N-N_0)} 
+ \epsilon_{1}{}_{_{\rm SR}}^{(-)}(N).
\label{eps1fit}
\end{equation}
Thus $\epsilon_1$ also decays exponentially during the transition,
from a very small value to another small value: in other words
$\epsilon_1$ is small even in the transition region around
$\phi_0$. These analytical estimates are compared to the exact
evolution of $\epsilon_1$ in Fig.~\ref{fig:eps1} and one again notices
that the matching is excellent.

\begin{figure*}
\includegraphics[width=0.45\textwidth,height=.45\textwidth]{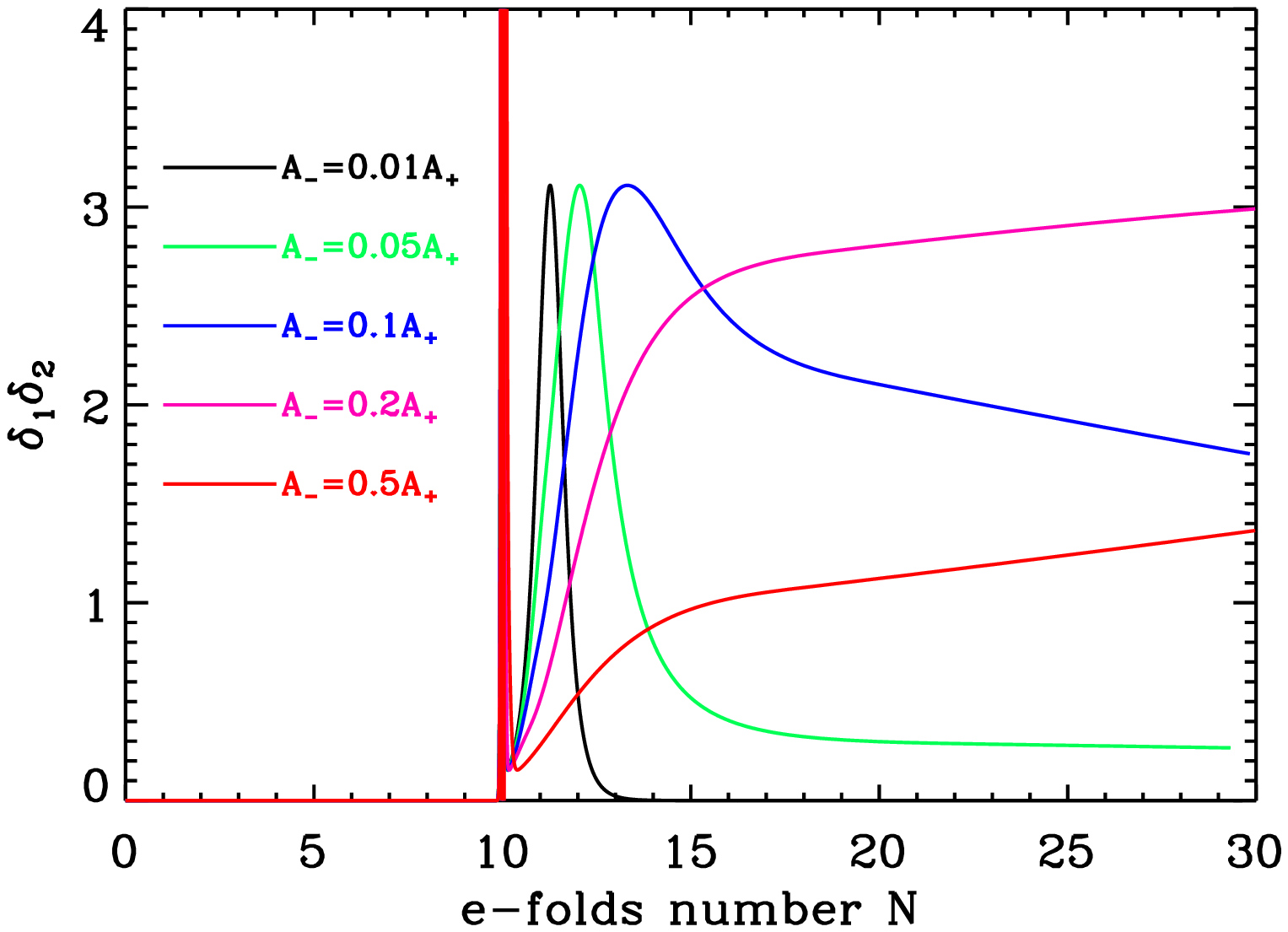}
\includegraphics[width=0.45\textwidth,height=.45\textwidth]{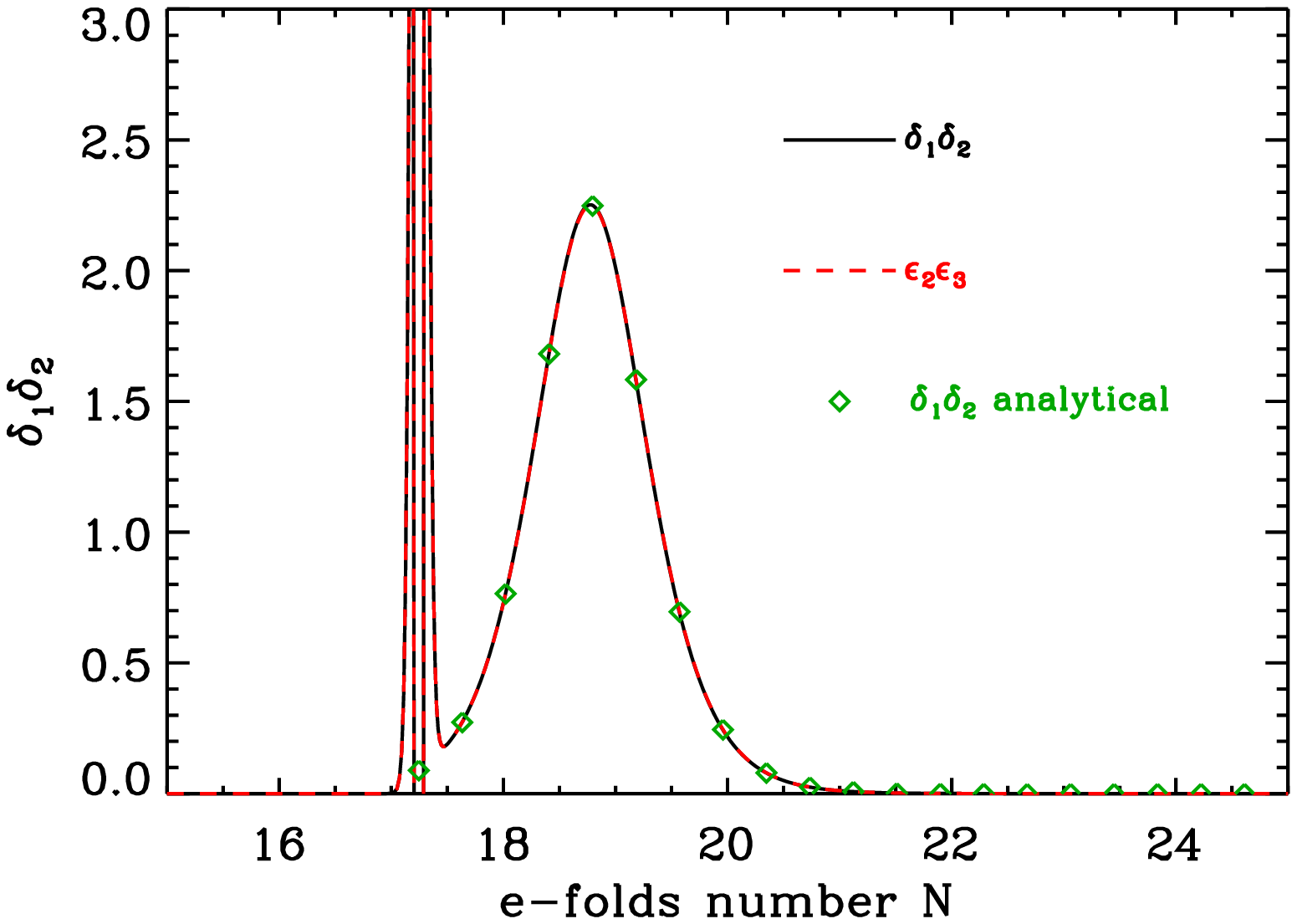}
\caption{{\bf Left panel}: evolution of $\delta_1\delta_2$ for "run
  one" (see the definition of ``run one" in the caption of
  Fig.~\ref{fig:field}) for $A_-=0.01 A_+$ (solid black line),
  $A_-=0.05 A_+$ (solid green line), $A_-=0.1 A_+$ (solid blue line),
  $A_-=0.2 A_+$ (solid pink line) and $A_-=0.5 A_+$ (solid red
  line). {\bf Right panel}: evolution of $\delta_1\delta_2$ (solid
  black line) and $\epsilon_2\epsilon_3$ (dashed red line) for ``run
  two" (see the definition of ``run two" in the caption of
  Fig.~\ref{fig:field}). We observe that $\delta _1\delta_2\simeq
  \epsilon_2\epsilon_3$. The dark green dots correspond to the
  ``slow-roll" solution given by Eq.~(\ref{eq:del12}) (and valid only
  after the transition). As can be noticed in the figure, this is an
  excellent fit to the numerical solution. The vertical lines at $N_0$
  originate from the Dirac function in $V_{\phi \phi}$ in
  Eq.~(\ref{bloodymess}).}
\label{fig:d1d2}
\end{figure*}

Let us now study the slow-roll parameters $\delta_1$ and
$\epsilon_2$. Upon using Eq.~(\ref{del1}), one obtains the following
expression valid only far from the transition
\begin{equation}
\delta_1{}_{_{\rm SR}}(N) 
\simeq 2 \sqrt{\frac{3}{\lambda V_0}} \phi(N).
\label{eq:deltasr}
\end{equation}
As opposed to $\epsilon_1$, this slow-roll parameter does not depend
on $A_\pm$. During the transition, one has
\begin{equation}
\delta_1^{(-)}(N) = \frac{ 3 \Delta A}{A_-}  
\frac{{\rm e}^{-3(N-N_0)} }{1-(\Delta A/A_-){\rm e}^{-3(N-N_0)}}+ 
\delta_{1}{}_{_{\rm SR}}(N).
\label{eq:delta1}
\end{equation}
The behaviour of the slow-roll parameter $\epsilon_2$ can be obtained
in the following way. Starting from Eq.~(\ref{VV}) it follows that
\begin{equation}
\Mp^2\frac{V_{\phi \phi}}{V}\simeq \frac{\gamma}{2} \left(4\epsilon_1
  -{\epsilon_2} -{\delta_1}\right).
\end{equation}
Thus in the slow-roll regimes on either side of $\phi_0$ where
$V_{\phi \phi}=0$, 
\begin{equation} 
4\epsilon_1{}_{_{\rm SR}}^{(\pm)}=
\epsilon_2{}_{_{\rm SR}}^{(\pm)} + \delta_1{}_{_{\rm SR}}.
\label{julien}
\end{equation}
On the other hand, we have established that during the transition,
$\epsilon_2\simeq \delta_1$ when $\gamma \gg 1$ [see
Eq.~(\ref{jumpd1})]. Thus on combining Eqs.~(\ref{eq:delta1}) and
(\ref{julien}), we find
\begin{equation}
\epsilon_{2}^{(-)}(N) = \frac{ 3 \Delta A}{A_-} \frac{{\rm e}^{-3(N-N_0)}
}{1-(\Delta A/A_-) {\rm e}^{-3(N-N_0)}}+4\epsilon_{1}{}_{_{\rm SR}}^{(-)}(N) -
\delta_{1}{}_{_{\rm SR}}(N),
\label{cool}
\end{equation}
where $\epsilon_{1}{}_{_{\rm SR}}^{(-)}$ and $\delta_{1}{}_{_{\rm
    SR}}$ are given in Eqs.~(\ref{eq:eps1SR}) and (\ref{eq:deltasr})
respectively. The above considerations are checked in
Fig.~\ref{fig:delta1} where the above analytical estimates are shown
to be excellent approximations to the exact numerical evolutions of
the two slow-roll parameters $\epsilon_2$ and $\delta _1$.

\par

Finally, we study the behaviour of the quantities $\delta _1 \delta_2
$ and $\epsilon_2 \epsilon_3$. These two combinations are important
because they appear in the effective potential for the cosmological
perturbations (see the next section). Just after the discontinuity,
where $\epsilon_2 \simeq \delta_1 \gg \epsilon_1$, we find from
Eqs.~(\ref{bloodymess}) and~(\ref{bloodymessbis}) that
\begin{eqnarray} 
\delta_{1} \delta_2 &\simeq& -3\delta_1 - \delta_1^2 - \frac{T_{\phi \phi}}{H^2} \; \;
\simeq\; \; -3\epsilon_2 - \epsilon_2^2 ,
\label{bm1}
\\
\epsilon_{2} \epsilon_3 &\simeq& -3\epsilon_2 - \epsilon_2^2,
\label{bm2}
\end{eqnarray}
where we have neglected the $T_{\phi \phi}$ term in $\delta_1
\delta_2$ which is small.  Thus, substituting (\ref{cool}) yields the
fit after the discontinuity of
\begin{equation}
\delta_1 \delta_2 \simeq \epsilon_2 \epsilon_3
\simeq -9 \frac{ \Delta A}{A_-}  \frac{{\rm e}^{-3(N-N_0)} }
{\left[ 1-(\Delta A/A_-) {\rm e}^{-3(N-N_0)}\right]^2}.
\label{eq:del12}
\end{equation}
This analytical approximation is tested in Fig.~\ref{fig:d1d2}. As can
be seen in this plot, the matching to the exact numerical solution is
excellent.

\par

To summarise, the above considerations show that on assuming that
potential is vacuum dominated, that the system is initially in
slow-roll, and that $\gamma \gg 1$ for all times, then it is possible
to obtain excellent analytical approximations for each relevant
background quantities and slow-roll parameters throughout the
evolution, despite the fact that slow-roll is violated at the
transition.  This is important since the slow-roll parameters control
the evolution of the time-dependent frequency of each Fourier mode of
the perturbations. In the next section we study in detail cosmological
fluctuations, our final goal being to determine their
power-spectrum. Before doing so, however, we end by noting that in the
slow-roll regime far from the discontinuity, $\gamma_{_{\rm
    SR}}^{(\pm)}$ as well as all the slow-roll parameters, only depend
very weakly on $N$: indeed they remain essentially constant on either
side of the discontinuity (as can be seen in Figs.~\ref{fig:field} -
\ref{fig:d1d2}). The origin of this behaviour is the small variation
of $\phi$ with $N$ as seen in Fig.~\ref{fig:field}. For this reason,
in the following, we will often omit to write the explicit $N$
dependence of the slow-roll parameters in the slow-roll regime. In the
above expressions, this amounts to approximating $\phi(N)$ by
$\phi(N_0)=\phi_0$, so that for example from (\ref{eq:eps1SR})
\begin{equation}
\epsilon_1{}_{_{\rm SR}}^{(\pm)}\simeq \frac{\Mp A_{\pm}}{2V_0}
\sqrt{\frac{3}{\lambda V_0}}\phi_0^2\, 
\end{equation} 
while from Eqs.~(\ref{gpma}) and~(\ref{eq:gamev})
\begin{equation} 
\gamma _{_{\rm SR}}^{(\pm)} \simeq 
\sqrt{\frac{\lambda}{3V_0}} \frac{\Mp A_{\pm}}{\phi_0^2} \, , \quad
 \gamma^{(-)}(N) \simeq \bgammam 
\left[ \frac{\Delta A}{A_-}{\rm e}^{-3(N-N_0)} - 1 \right]\, .
\label{eq:gamevb}
\end{equation}
Having completed the study of the background, we now turn to the
perturbations, in particular to the calculation of the two-point
correlation function.

\section{Power-spectrum}
\label{sec:powerspectrum}

In this section, we are interested in scalar perturbations. It is
well-known that they can be characterised by a single variable, the
so-called Mukhanov-Sasaki quantity, $v(\eta,{\bm x})$. Its Fourier
amplitude obeys the equation of a parametric oscillator, namely
\begin{equation}
v_{\bm k}''+\left(\frac{k^{2}}{\gamma^2}
-\frac{z''}{z}\right)v_{\bm k}=0 ,
\label{eq:eomv}
\end{equation}
where the prime now denotes a derivative with respect to conformal
time, $k$ denotes the comoving wave number of the Fourier mode under
consideration, and $z$ is given by \cite{Lorenz:2008et}
\begin{equation}
z(\eta) = a(\eta) \Mp  \sqrt{2\epsilon_1} \gamma\, .
\label{zdef}
\end{equation}
The effective potential $z''/z$ which determines the evolution of the
scalar perturbations can be determined directly from Eq.~(\ref{zdef})
and the definition of the slow-roll parameters in Eqs.~(\ref{epsdef})
and (\ref{deldef}): its exact expression is
\begin{equation}
  \frac{z''}{z} =a^2 H^2\left[2-\epsilon_1 + \frac{3}{2}\epsilon_2 
+ \frac{1}{4}\epsilon_2^2 - \frac{1}{2}\epsilon_1\epsilon_2  
+ \frac{1}{2}\epsilon_2\epsilon_3  
+ \left(3-\epsilon_1+\epsilon_2\right)\delta_1 + \delta_1^2 
+ \delta_1 \delta_2 \right].
\label{eq:bmess}
\end{equation}
(Note that if $\delta_1=\delta_2=0$, this reduces to the SSFI
expression, see eg.~\cite{Martin:2011sn}). Another important
difference with respect to conventional inflationary theory is the
presence of the term $1/\gamma^2 =c_{_{\rm S}}^2$ in front of $k^2$ in
Eq.~(\ref{eq:eomv}), responsible for the fact that perturbations now
propagate with a speed $c_{_{\rm S}}$ different to the speed of light.
As a result, the question of how initial conditions are chosen is more
subtle --- indeed, one usually assumes that the initial state is the
adiabatic vacuum for which
\begin{equation}
v_{\bm k}(\eta)\simeq \frac{1}{\sqrt{2\omega(k,\eta)}}
\exp\left[\pm i\int ^{\eta }\omega(k, \tau){\rm d}\tau\right],
\end{equation}
where $\omega^2(k,\eta)\equiv c_{_{\rm S}}^2k^2-z''/z$. The adiabatic
approximation is valid if $\vert Q/\omega ^2\vert\ll 1$ where $Q\equiv
3\omega'^2/(4\omega^2)-\omega''/(2\omega)$. In the standard
inflationary context, the modes are initially within the Hubble radius
and $\omega \simeq k$. It is then obvious that the
Wentzel-Kramers-Brillouin (WKB) approximation is valid. In the DBI
case, however, the modes are initially within the sonic scale and
$\omega \simeq c_{_{\rm S}}k$. Because of the non-trivial
time-dependence of $c_{_{\rm S}}$, it is not obvious that $\vert
Q/\omega^2\vert \ll 1$. If this is the case, then one simply looses
the ability to choose a well-defined and well-motivated initial
state. In fact, it is easy to show that, on sub-sonic scales,
\begin{equation}
\frac{Q}{\omega^2}=\frac{a^2H^2}{2c_{_{\rm S}}^2k^2}
\left(\delta_1-\epsilon_1\delta_1+\delta_1\delta_2+\frac12 \delta_1^2\right).
\end{equation}
We have seen that, in the DBI-Starobinsky model, all the slow-roll
parameters are small initially (\ie far from the transition). It
follows that, in this model, one can identify a well-defined initial
state $v_{\bm k}(\eta)={{\gamma}/{(2k})}\, {\rm e}^{\pm {
    ik\eta}/{\gamma}}$. This choice is made in the remainder of this
article.

\par

\begin{figure}
\includegraphics{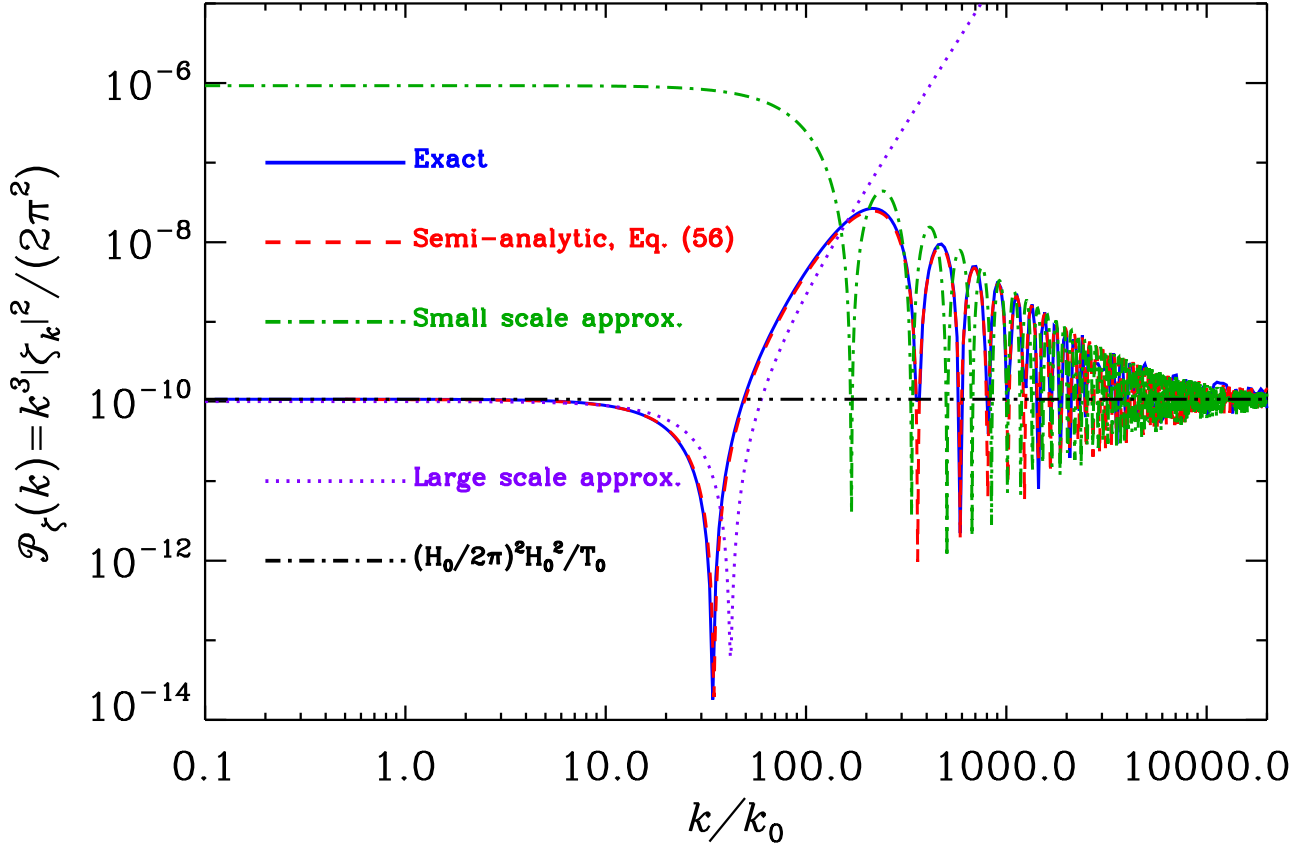}
\caption{Power spectrum of the DBI-Starobinsky model and various
  approximations. The solid blue line represents the exact power
  spectrum obtained with a numerical, mode by mode, integration for
  ``run two''. The dashed red line gives ${\cal P}_\zeta$ obtained
  from an integration of Eq.~(\ref{theequation}). The dotted dashed
  green line comes from Eq.~(\ref{eq:Plargek}) and is relevant only in
  the small scale limit. The purple dotted line represents the
  approximation~(\ref{pk2}) and is valid on large scales only.
  Finally, the black dotted-dotted-dashed horizontal line gives the
  asymptotic values of the power spectrum, see
  Eqs.~(\ref{eq:verylargescale}) and~(\ref{pk1}).}
\label{fig3}
\end{figure}

The power-spectrum, or two-point correlation function, is defined by
\begin{equation} 
{\cal P}_{\zeta}\equiv \frac{k^3}{2\pi^2}\vert
  \zeta_{\bm k}\vert ^2 =\frac{k^3}{4\pi^2}\frac{c_{_{\rm S}}^2\vert
    v_{\bm k}\vert^2}{\Mp^2a^2\epsilon_1},
\label{Pzeta}
\end{equation}
where $\zeta_{\bm k}\equiv v_{\bm k}/z$ is the curvature perturbation,
and the right hand side is evaluated in the limit, $k\cs/(aH) \simeq
-k\eta /\gamma \rightarrow 0$. To find ${\cal P}_\zeta(k)$, we must
integrate (\ref{eq:eomv}) for each mode starting from the initial
conditions discussed above. In general this cannot be done
analytically due to the complexity of the equations. As a result, one
possibility is to integrate the system numerically, and for this
purpose we have written a numerical code which exactly integrates the
background and the perturbations mode by mode (and for arbitrary
values of the parameters of the model, and hence any $\cs$).  The
result is displayed in Fig.~\ref{fig3} (solid blue line). We see that,
as $k$ increases, the power-spectrum rapidly dips and reaches its
minimum at a scale $k_*$ which is a large multiple of $k_0 \equiv
-1/\eta_0$ (namely the mode which left the Hubble radius at
$\phi=\phi_0$).\footnote{Notice that we have chosen to define $k_0$
  without any factor of $\gamma$ despite the fact that, physically,
  the relevant scale is $aH/\cs$. The reason for this is that, in this
  model, $\gamma$ varies significantly with time and it is not very
  convenient to include it in the definition of the preferred scale.}
This dip is followed by large amplitude, high frequency
oscillations. Notice that for the parameters chosen in
Fig.~\ref{fig3}, the power-spectrum takes the same scale-invariant
value on large and small scales.  In the conclusion, we will compare
this power-spectrum with that obtained in the canonical Starobinsky
model for precisely the same values of $A_\pm$: as we will see,
although the shapes share the same general aspect, they are in fact
very different.

It is also interesting to have an analytical expression for the power
spectrum since this can help understand how its shape is modified when
the parameters of the model are changed. Therefore we now aim to
develop approximation methods in order to address this question. To
solve Eq.~(\ref{eq:eomv}) and hence determine the $v_{\bm k}$ and
${\cal P}_\zeta(k)$, we use the results of section
\ref{subsec:application} in which the slow-roll parameters were
determined as a function of $N$, and hence as a function of conformal
time $\eta$ since ${\eta} = {\eta_0}{\rm e}^{-(N-N_0)}$. Our
analytical approximation will therefore only be valid in the $\cs \ll
1$ regime for which the results of section \ref{subsec:application}
hold.  Before the transition $\eta < \eta_0$, the slow-roll parameters
are small and Eq.~(\ref{eq:bmess}) reduces to $z''/z \simeq 2 a^2 H^2
\simeq 2/\eta^2 $ so that the solution of (\ref{eq:eomv}) is
\begin{equation}
v_{\bm k}^+(\eta)  = \sqrt{\frac{\bgammap}{2k}}
\left( 1 - i \frac{\bgammap}{k \eta} \right){\rm e}^{-ik\eta/\bgammap},
\label{viper}
\end{equation}
where $\bgammap$ is given in Eq.~(\ref{eq:gamevb}). After the
transition, the effective potential $z''/z$ can be calculated using
the fact that $\delta_1 \simeq \epsilon_2 \gg \epsilon_1$ while
$\delta_1 \delta_2$ and $\epsilon_1 \epsilon_2$ are given in
Eq.~(\ref{bm1}) and (\ref{bm2}) respectively.  On substituting, one
finds that the terms linear in slow-roll parameters in
(\ref{eq:bmess}) cancel (just as in the canonical Starobinsky model),
but the quadratic terms do not, resulting in ${z''}/{z} \simeq {\cal
  H}^2 (2 + 3\epsilon_2^2/4)$. Thus after the transition, the mode
function satisfies
\begin{equation}
v^-_{\bm k}{}'' + \left\{ \frac{k^{2}}{\gamma^2(\eta)} 
- \frac{1}{\eta^2}\left[2  + \frac{3}{4}\epsilon_2^2(\eta) \right] 
\right\}v_{\bm k}^- = 0\, ,
\label{step1}
\end{equation}
where, from Eq.~(\ref{cool})
\begin{equation}
  \epsilon_2 = \epsilon_{2}^{(-)} \simeq \frac{ 3 \Delta A}{A_-} 
\frac{{\rm e}^{-3(N-N_0)}}{1-(\Delta A/A_-) {\rm e}^{-3(N-N_0)}} 
=  3  \frac{\tilde{\omega} \eta^3}{(1-\tilde{\omega} \eta^3)}
\end{equation}
with $\tilde{\omega} \equiv (\Delta A/A_-)/\eta_0^3 >0$. Thus, on
using Eq.~(\ref{eq:gamevb}), we finally arrive at
\begin{equation}
  v^-_{\bm k}{}'' + \left[\frac{k^2}{\left(\bgammam\right)^2 (1-\tilde{\omega}
      \eta^3)^2} - \frac{2}{\eta^2} - \frac{27}{4}
    \frac{\tilde{\omega}^2 \eta^4}{(1-\tilde{\omega} \eta^3)^2}
  \right]v^-_{\bm k}=0 \, .
\label{theequation}
\end{equation}
This equation is one of the central results of this paper, since the
power-spectrum is determined directly from its solution. In order to
find ${\cal P}_\zeta(k)$, the modes $v^\pm_{\bm k}$ and their
derivatives must be matched at $\eta=\eta_0=-1/(a_0 H_0)$ [so as to
determine the two integration constants associated with
Eq.~(\ref{theequation})]. This can be done by carefully considering of
the behaviour of $z$ and $z''/z$ across the transition. Indeed, on
recalling that $z=M_{\rm Pl} a(\eta) \sqrt{2\epsilon_1(\eta)}
\gamma(\eta)$, and using Eqs.~(\ref{eq:gamev}) and (\ref{eps1fit}),
one finds
\begin{equation}
\frac{z''}{z} \simeq - \frac{9}{2} \frac{\Delta A}{A_+} 
\frac{1}{\eta_0} \, \delta^{(1)}(\eta - \eta_0) =
 a_0 H_0 \frac{9}{2} \frac{\Delta A}{A_+}\, \delta^{(1)}(\eta - \eta_0).
\end{equation}
Hence the matching conditions at the transition are
\begin{equation}
v^-_{\bm k}(\eta_0) = v^+_{\bm k}(\eta_0)
\, , \quad
v'^-_{\bm k}(\eta_0) - v'^+_{\bm k}(\eta_0) 
= a_0 H_0 \frac{9}{2} \frac{\Delta A}{A_+} v^-_{\bm k}(\eta_0).
\label{match2}
\end{equation}
Unfortunately, Eq.~(\ref{theequation}) is not soluble
analytically. However, it can be solved numerically. At this point,
one could wonder whether we have gained something given that our aim
was to derive approximate analytical formulae and that we have already
determined ${\cal P}_\zeta$ exactly by means of a mode by mode
integration. However, integrating a single differential equation is
much easier than writing a mode by mode numerical code and, moreover,
as we will discuss below, Eq.~(\ref{theequation}) can also be
approximated analytically. In Fig.~\ref{fig3} we plot the
power-spectrum obtained by solving (\ref{theequation}) numerically,
and then substituting into (\ref{Pzeta}) (red dashed curve). This is
compared with the fully mode by mode numerical calculation of the
power-spectrum (solid blue line) obtained by solving both the
background and perturbation equations numerically. The agreement is
excellent, thus validating all the reasoning used to arrive at
(\ref{theequation}).

\begin{figure}
\includegraphics[width=0.85\textwidth,height=.65\textwidth]{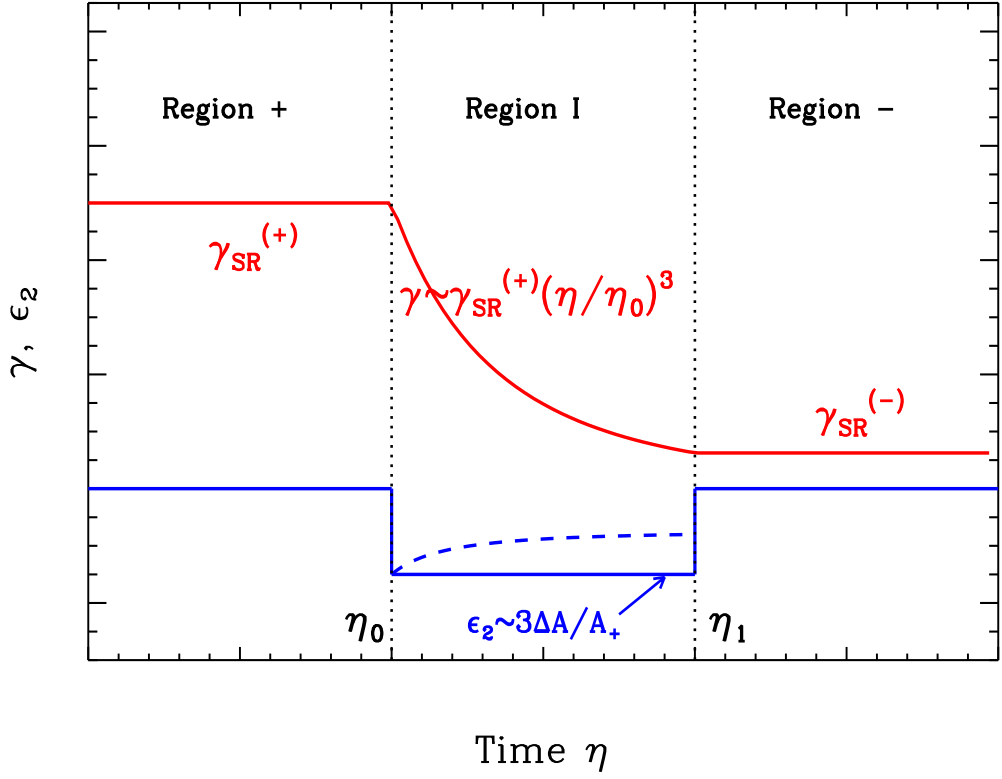}
\caption{Approximations for an analytic solution of the mode equation
  (\ref{theequation}). Solid lines: approximations for $\gamma$ (red)
  and $\epsilon_2$ (blue) made on small scales. On large scales, the
  $\eta$-dependence of $\epsilon_2$ is taken into account (dashed
  line), namely from Eq.~(\ref{cool}), $\epsilon_2\simeq (3 \Delta
  A/A_-){\rm e}^{-3(N-N_0)} /\left[1-(\Delta A/A_-) {\rm
      e}^{-3(N-N_0)}\right]$.}
\label{fig:matching}
\end{figure}

Our aim is now to determine analytically the dependence of the
features described above on the parameters of the model $A_\pm$. On
large scales, the behaviour of ${\cal P}_\zeta(k)$ can be captured via
the following approximation scheme.  Since $\gamma \gg 1$, on large
scales $k\eta_0/\bgammam \ll 1$, we neglect the first term in
Eq.~(\ref{theequation}) while keeping the exact $\eta$-dependence of
$\epsilon_2$, see Fig.~\ref{fig:matching}. It then follows that
\begin{equation} 
  v^{-}_{{\bm k}, {\rm large}}= \frac{1}{\sqrt{1-\tilde{\omega}
      \eta^3}} \left[ \frac{\tilde{C}_3(k) }{\eta} + \tilde{C}_4(k)
  \eta^2 (-2+\tilde{\omega} \eta^3)\right]\, ,
\label{vlarge}
\end{equation}
where the $k$-dependent integration constants $\tilde{C}_{3,4}(k)$ can be
determined straightforwardly from the boundary conditions at $\eta_0$,
given in Eq.~(\ref{match2}) --- in particular
\begin{equation} 
\tilde{C}_3(k) = \frac{i}{\sqrt{2}} \left(\frac{\bgammam}{k}\right)^{3/2}
\left[- 1 -i \frac{ k \eta_0}{ \bgammap}
+ \left(\frac{k \eta_0}{\bgammap }\right)^2 
\frac{A_+({A_-}+{A_+})}{6A_-^2}  \right]\, .
\label{C3t}
\end{equation}
From Eq.~(\ref{Pzeta}) we then have 
\begin{eqnarray}
\lim_{k/(k_0 \bgammam) \ll 1} {\cal P}_\zeta(k) &=&
 \left(\frac{H_0}{2\pi} \right)^2 \left( \frac{H_0^2}{T_0} \right) \left[ 
\frac{2 k^3 |\tilde{C}_3(k)|^2}{(\bgammam)^3} \right],
\label{pk2}
\end{eqnarray}
where $T_0 =T(\phi_0)$. This expression is drawn by a purple dotted
line in Fig.~\ref{fig3} and we see that, on large scales, the matching
is good and the first dip is predicted with a reasonable precision. In
the limit $k/k_0\rightarrow 0$, one has 
\begin{equation}
  \lim_{k/k_0 \rightarrow 0} {\cal P}_\zeta(k) =
  \left(\frac{H_0}{2\pi}\right)^2\left(\frac{H_0^2}{T_0}\right)
\label{eq:verylargescale} 
\end{equation}
(represented by the black dotted-dotted-dashed line in
Fig.~\ref{fig3}), which is indeed an accurate prediction of the
overall amplitude of the spectrum on large scales. From
Eq.~(\ref{C3t}) we can also estimate the smallest value of $k=k_*$ for
which ${\cal P}_\zeta(k_*)\rightarrow 0$: we find $k_*/k_0 \simeq
\bgammap A_- [6/(A_+(A_-+A_+))]^{1/2}$ which, for the parameters of
Fig.~\ref{fig3} gives $k_*/k_0 \sim 42$ thus agreeing quite well with
the numerical value.\footnote{A more accurate expression for $k_*$
  could be obtained by solving Eq.~(\ref{theequation}) perturbatively
  in $k$, using (\ref{vlarge}) as the zeroth order solution.}

On small scales we proceed as follows. First introduce a conformal
time $\eta_1 > \eta_0$ so that the period after the transition is
split into two parts, see Fig.~\ref{fig:matching}. When $\eta \geq
\eta_1$, which we denote by `region $-$', we take $\epsilon_2\simeq 0$
and $\gamma=\bgammam$. The solution of Eq.~(\ref{step1}) is then
particularly straightforward, namely
\begin{equation}
v_{{\bm k}}^- = \alpha_{{\bm k}}  \sqrt{ \frac{\bgammam}{2k}} 
\left(1-i\frac{\bgammam}{k\eta} \right){\rm e}^{-ik\eta/\bgammam} + 
\beta_{ \bm k}   \sqrt{ \frac{\bgammam}{2k}} 
\left(1+i\frac{\bgammam}{k\eta} \right){\rm e}^{ik\eta/\bgammam}\, .
\label{vminus}
\end{equation}
The integration constants $\alpha_{ \bm k}$ and $\beta_{ \bm k}$, to
be calculated below, determine the power-spectrum since from
Eq.~(\ref{Pzeta}) it follows that
\begin{equation}
{\cal P}_\zeta(k) =  \frac{H_-^2}{8\pi^2 \Mp^2}  
\frac{ \bgammam}{\epsilon_1^-} |\beta_{ \bm k} - \alpha_{\bm k}|^2\, = 
\left(\frac{H_0}{2\pi} \right)^2 \left( \frac{H_0^2}{T_0} \right)  
\left|\beta_{ \bm k} - \alpha_{ \bm k}\right|^2.
\label{Pin}
\end{equation}
In the intermediate region, ``region I'' for
which $\eta_0 \leq \eta < \eta_1$, we make the following assumptions
(see Fig.~\ref{fig:matching}): $\epsilon_2$ is taken constant, with
$\epsilon_2 \simeq \epsilon_2(\eta_0) = 3\Delta A/A_+ $ as obtained
from Eq.~(\ref{cool}), while from Eq.~(\ref{eq:gamev}) the evolution
of $\gamma(\eta)$ is approximated by $\gamma\simeq \bgammap
\left(\eta/\eta_0\right)^3 $. Continuity of $\gamma$ at $\eta_1$
determines $\eta_1$ to be given by $\eta_1=\eta_0
({\bgammam}/{\bgammap})^{1/3}$.  In that case, Eq.~(\ref{step1}) has
the exact solution
\begin{equation} 
  v^{\rm I}_{{\bm k}, {\rm small}}=\sqrt{\frac{\eta}{\eta_1}}\left[C_1(k) J_\nu(x) +
   C_2(k) Y_\nu(x) \right]
\label{vsmall}
\end{equation}
where $C_1(k), C_2(k)$ are $k$-dependent integration constants, and
\begin{equation}
  x(\eta,k)\equiv \frac{1}{2}\left( \frac{k\eta_0}{\bgammap} \right) 
  \left(\frac{\eta_0}{\eta}\right)^2 \, 
  \, , \qquad \nu \equiv  \frac{3}{4} 
\sqrt{ 1 + 3 \left(\frac{\Delta A}{A_+}\right)^2 }\, .
 \end{equation}
The integration constants $C_{1,2}(k)$ are determined from the
boundary conditions at $\eta_0$ which, from Eq.~(\ref{match2}), read
\begin{eqnarray}
v^{\rm I}_{\bm k}(\eta_0) &=& v^+_{\bm k}(\eta_0), \quad
v^{\rm I}_{\bm k}{}'(\eta_0) - v^+_{\bm k}{}'(\eta_0) 
=a_0 H_0 \frac{9}{2} \frac{\Delta A}{A_+} v^+_{\bm k}(\eta_0),
\end{eqnarray}
where the mode solutions in the ``+" region, $v_{\bm k}^+$, are given
in Eq.~(\ref{viper}). This leads to the following expressions
\begin{eqnarray}
C_1 &=& \frac{\pi}{2}\sqrt{\frac{\bgammam}{2k}}\left(x_0x_1\right)^{1/2}
{\rm e}^{-2ix_0}\left\{-i
\left[1+\frac{3}{2\left(2ix_0\right)^2}\left(1+2ix_0\right)
\left(1+3\frac{\Delta A}{A_+}\right)\right]Y_{\nu}\left(x_0\right)
+\frac{1+2ix_0}{2ix_0}Y_{\nu}'\left(x_0\right)\right\}, \\
C_2 &=& \frac{\pi}{2}\sqrt{\frac{\bgammam}{2k}}\left(x_0x_1\right)^{1/2}
{\rm e}^{-2ix_0}\left\{i
\left[1+\frac{3}{2(2ix_0)^2}\left(1+2ix_0\right)
\left(1+3\frac{\Delta A}{A_+}\right)\right]J_{\nu}\left(x_0\right)
-\frac{1+2ix_0}{2ix_0}J_{\nu}'\left(x_0\right)\right\},
\end{eqnarray}
where we have used the notation $x_0\equiv
x(\eta_0,k)=k\eta_0/(2\bgammap)$. In turn, the integration constants
$\alpha_{ \bm k}$ and $\beta_{ \bm k}$ are then obtained from matching
at $\eta_1$, where the relevant conditions are $v_{\bm k}^{\rm
  I}(\eta_1) =v_{\bm k}^{-}(\eta_1)$ and $v_{\bm k}^{\rm I}{}'(\eta_1)
=v_{\bm k}^{-}{}'(\eta_1)$, namely the mode functions and their
derivatives are continuous. One obtains
\begin{eqnarray}
\alpha_{\bm k} &=& -\sqrt{\frac{2k}{\bgammam}}{\rm e}^{2ix_1}\left\{
\frac{3-6ix_1-8x_1^2}{16x_1^2}
\left[C_1J_{\nu }(x_1)+C_2Y_{\nu}(x_1)\right]
-
\frac{1-2ix_1}{4x_1}
\left[C_1J_{\nu }'(x_1)+C_2Y_{\nu}'(x_1)\right]\right\} 
\\
\beta_{\bm k} &=& -\sqrt{\frac{2k}{\bgammam}}{\rm e}^{-2ix_1}
\left\{
\frac{3+6ix_1-8x_1^2}{16x_1^2}
\left[C_1J_{\nu }(x_1)+C_2Y_{\nu}(x_1)\right]
-
\frac{1+2ix_1}{4x_1}
\left[C_1J_{\nu }'(x_1)+C_2Y_{\nu}'(x_1)\right]\right\}, 
\end{eqnarray}
where, this time, we have introduced the notation $x_1\equiv
x(\eta_1,k)=k\eta_1/(2\bgammam)$. Since we are interested in deriving
the expression of the power spectrum on small scales, one can take the
limit $x_1\rightarrow \infty$ in the two above expressions. In that
case, they reduce to
\begin{eqnarray}
\alpha _{\bm k} &\rightarrow & \frac{1}{2}\sqrt{\frac{2k}{\bgammam}}
\sqrt{\frac{2}{\pi x_1}}{\rm e}^{i\left(3x_1-\pi\nu /2-\pi/4\right)}
\left(C_1-iC_2\right), \\
\beta _{\bm k} &\rightarrow & \frac{1}{2}\sqrt{\frac{2k}{\bgammam}}
\sqrt{\frac{2}{\pi x_1}}{\rm e}^{-i\left(3x_1-\pi\nu /2-\pi/4\right)}
\left(C_1+iC_2\right).
\end{eqnarray}
To go further, we see that we now need the combinations $C_1\pm
iC_2$. Moreover, one can use the fact that the parameter
$\epsilon\equiv A_-/A_+$ is small. Indeed, if this is not the case,
then the exact numerical integration indicates that the amplitude of
the oscillations becomes large and the model becomes obviously ruled
out. One can therefore expand $C_1$ and $C_2$ in $\epsilon\equiv
A_-/A_+$, which in fact amounts to considering that $\nu=3/2$ in the
formulae giving $C_1$ and $C_2$. One obtains
\begin{eqnarray}
C_1+iC_2 &=& \frac{3}{4x_0}\sqrt{\frac{\pi \bgammam}{k}}x_1^{1/2}
{\rm e}^{-ix_0+i\pi/2}\left[1+{\cal O}\left(\epsilon\right)\right],\\
C_1-iC_2 &=& \frac{1}{4x_0}\sqrt{\frac{\pi \bgammam}{k}}x_1^{1/2}
\left(-4x_0+3i\right){\rm e}^{-3ix_0}
\left[1+{\cal O}\left(\epsilon\right)\right].
\end{eqnarray}
Let us stress that, in the two above equations, no limit
$x_0\rightarrow \infty$ has been taken and that the result is
valid for any $x_0$. Finally, straightforward manipulations leads to
the following expression
\begin{equation}
\left \vert \beta _{\bm k}-\alpha _{\bm k}\right\vert ^2
\simeq 1+\frac{3}{2x_0}\sin\left(6x_1-2x_0\right)+\frac{9}{4 x_0^2}
\sin^2 \left(3x_1-x_0\right),
\end{equation}
from which the power-spectrum ${\cal P}_\zeta(k)$ is then obtained
from Eq.~(\ref{Pin}). Our analytic result then gives
\begin{equation}
\label{eq:Plargek}
{\cal P}_\zeta^{(A_- \ll A_+)}(k) =  \left( \frac{H_0}{2\pi} \right)^2 
\left(\frac{H_0^2}{T_0} \right) 
\left\{ 1 +   \left(\frac{ 3 \bgammap}{k\eta_0}\right)  \sin [2 \theta(k) ]   
+  \left(\frac{ 3 \bgammap}{k\eta_0}\right)^2\sin^2[\theta(k)] \right\} 
\qquad (\text{for} \; \; k\gg k_0)
\end{equation}
where, using the explicit expressions of $x_0$ and $x_1$, the argument
$\theta(k)$ of the trigonometric functions in the above formula can be
written as
\begin{equation}
  \theta(k) = \frac{1}{2}\left( \frac{k\eta_0}{\bgammap} \right) 
  \left[ 3 \left(\frac{\bgammap}{\bgammam}\right)^{2/3} - 1 \right] 
  \simeq \frac{3}{2}\left( \frac{k\eta_0}{\bgammap} \right) 
  \left(\frac{\bgammap}{\bgammam}\right)^{2/3}  \quad {\rm for} \quad
  \bgammap \gg \bgammam.
\end{equation}
Thus the wavelength of the oscillations is approximately given by
$\sim \bgammap \left(\bgammam/\bgammap\right)^{2/3}$ and depends on
$A_\pm$ through the dependence of $\gamma _{_{\rm SR}}^{(\pm)}$ on
these parameters. Relative to the asymptotic value of ${\cal
  P}_\zeta(k\rightarrow \infty)$, the amplitude of these oscillations
is determined by the ratio $3 \bgammap k_0/k$. When $3 \bgammap
k_0/k>1$, it is quadratic in this parameter, while when $3 \bgammap
k_0/k<1$, it is linear. This can be observed in Fig.~\ref{fig3} where,
for $k/k_0\simeq 3 \bgammap\simeq 5\times 10^3$, the slope of the
envelope can be seen to change. The expression~(\ref{eq:Plargek}) is a
good fit to the numerical result (see Fig~\ref{fig3}, dotted dashed
green line). We can also use it to extract the $k\rightarrow \infty$
behaviour of the power-spectrum. Indeed, we find that the scale invariant
value of the power-spectrum on small scales is given by
\begin{equation}
  \lim_{k/k_0 \rightarrow \infty} {\cal P}_\zeta(k) =
  \left( \frac{H_0}{2\pi} \right)^2 \left(\frac{H_0^2}{T_0} \right) 
 \label{pk1}
\end{equation}
as advertised in the introduction, and also shown in
Fig.~\ref{fig3}. We have thus proved that the overall amplitude on
large and small scales of the spectrum is the same. This is a peculiar
feature of the DBI Starobinsky model which makes it very different
from the standard canonical Starobinsky model.

\section{Discussion and Conclusions}
\label{sec:conclusions}

The main purpose of this paper was to study the signatures of the
inflationary DBI-Starobinsky model, for which the potential is linear
with a sharp change in slope at a certain $\phi_0$ and the kinetic
term a non-minimal, DBI, one. In the case of canonical inflation with
such a potential, both the power-spectrum ${\cal P}_\zeta(k)$ as well
as the bi-spectrum are exactly soluble
analytically~\cite{Martin:2011sn,Arroja:2012ae}. Here we have addressed the
following questions: what is the shape of ${\cal P}_\zeta(k)$ in the
DBI-Starobinsky case? What signature do the non-linear kinetic terms
leave? Does ${\cal P}_\zeta(k)$ still rise sharply from small to large
scales?

\begin{figure*}
\includegraphics[width=0.85\textwidth,height=.65\textwidth]{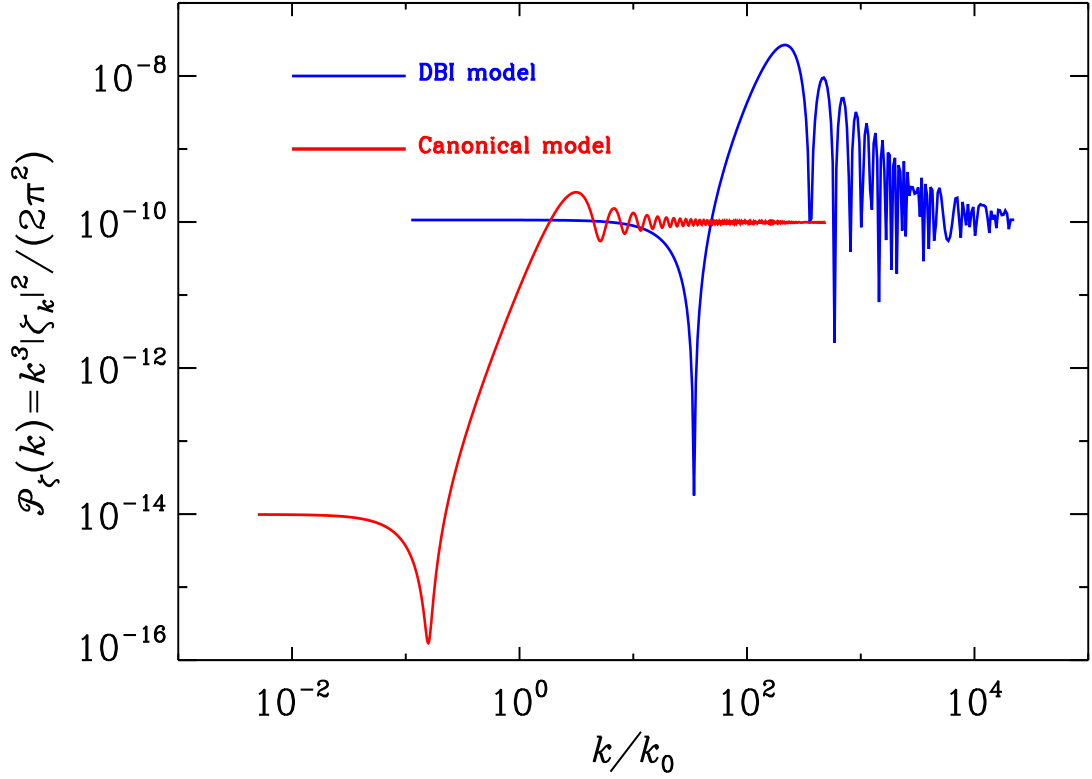}
\caption{Comparison of the DBI-Starobinsky and CS-Starobinsky power
  spectra for the same values of $A_+$ and $A_-$ corresponding to
  ``run two''. Obviously, the amplitude of the oscillations is much
  smaller in the CS-Starobinsky case but there is a rise in power that
  does not exist in the DBI-Starobinsky power spectrum.}
\label{temp-dani}
\end{figure*}

To approach this problem, first we studied the homogeneous background
evolution of the field and generalised slow-roll parameters. We showed
in section \ref{subsec:application} that in the DBI regime $\gamma \gg
1$, these quantities can all be determined analytically to very high
accuracy.  We also showed that this model is characterised by a first
slow-roll parameter $\epsilon_1$ which is tiny throughout the
evolution of the system, while the other generalised slow-roll
parameters all become large after the transition at $\phi_0$, decaying
back to small values over a few e-folds. Armed with the analytical
expressions for the slow-roll parameters, we showed that the
power-spectrum can simply be obtained by solving the mode equation
(\ref{theequation}). Furthermore, a numerical solution of this
equation was shown to agree exactly with a fully numerical
determination of the power-spectrum of the model (both at the
back-ground and perturbative level), see Fig.~\ref{fig3}.

It is also interesting to compare our results to the predictions of
the canonical Starobinsky model, see Fig.~\ref{temp-dani}. Following
an analytical approximation of Eq.~(\ref{theequation}), we were able
to show that in the DBI Starobinsky model, it is actually the second
dimensionful potential $T(\phi)$ (rather than $A_\pm$) which
determines the power-spectrum on small and large scales, as summarised
in Eqs.~(\ref{pk2}) and (\ref{pk1}). Thus as opposed to the canonical
Starobinsky model in which there is a sharp rise in power across $k_0$
if $A_- \ll A_+$, in the DBI-Starobinsky model there is {\it no} rise
in power, see Fig.~\ref{temp-dani}. Finally, in the $A_- \ll A_+$
limit, we have shown that the wavelength of oscillations in the
power-spectrum does depend on $A_\pm$, as opposed to the CS model, see
Fig.~\ref{temp-dani}. Furthermore, relative to the asymptotic value of
${\cal P}_\zeta(k\rightarrow \infty)$, the amplitude of oscillations
is now much larger --- rather than being of order $|\Delta A/A_+| \sim
1$ (in the CS model), it is now of order $\bgammap \gg 1$.

\begin{figure*}
\includegraphics[width=0.85\textwidth,height=.65\textwidth]{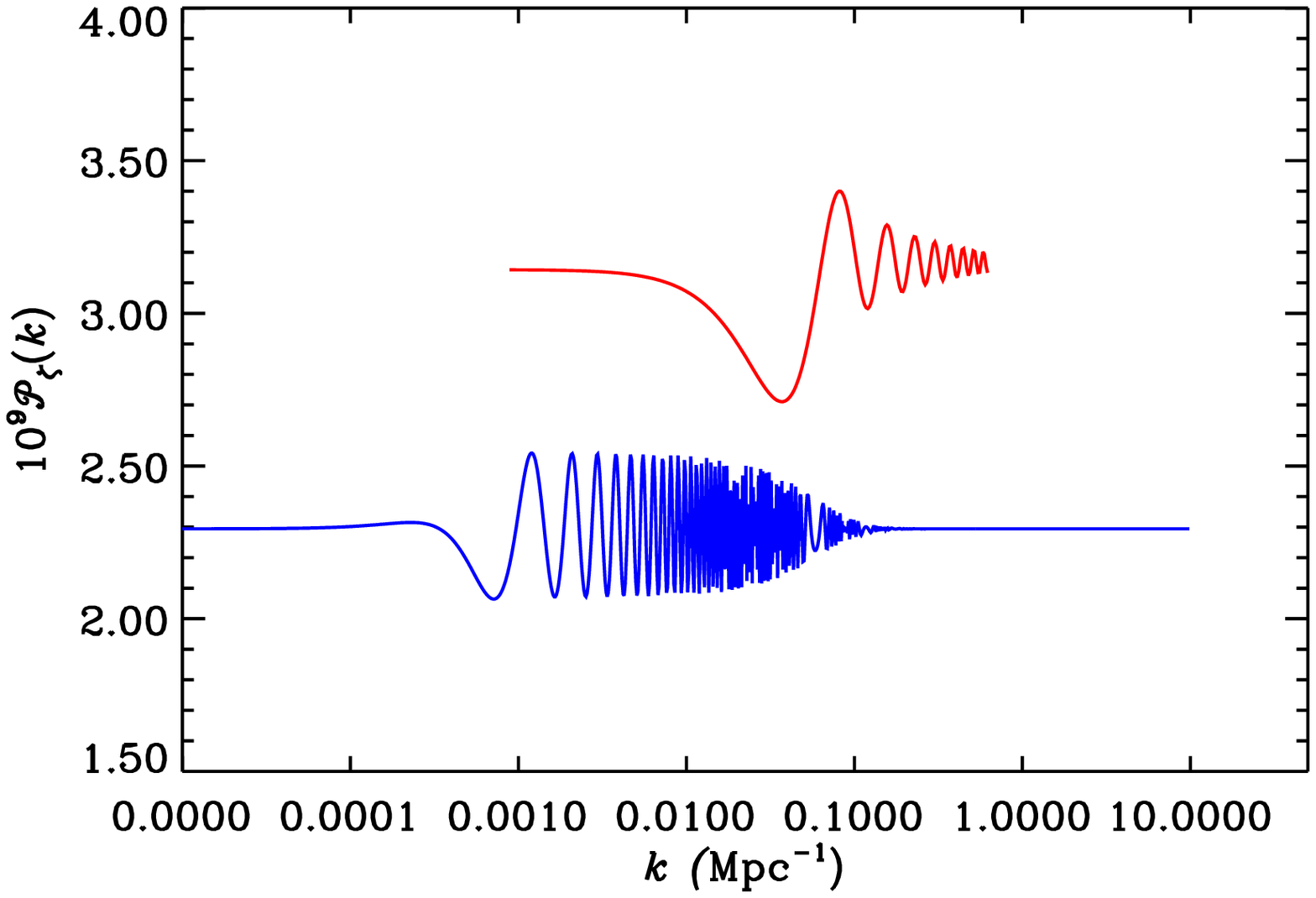}
\caption{Comparison of the DBI-Starobinsky power spectrum (solid red
  line) with the Planck ``step-inflation'' model best fit (solid blue
  line). It should be noticed that the DBI-Starobinsky power spectrum
  normalisation is arbitrary. The DBI-Starobinsky model corresponds to
  $\epsilon_{1\uin}\simeq 10^{-5}$, $\epsilon_{2\uin}\simeq 2.5\times
  10^{-5}$, $\delta _{1\uin}\simeq 1.\times 10^{-4}$, $\gamma
  _{\uin}\simeq 5$, $\phi_\uin/\Mp\simeq 77.4193$, $H_\uin/\Mp \simeq
  8.88\times 10^{-7}$ and $N_0=10$, where $N_0$ is the number of
  e-folds at which the field goes through the transition. This implies
  $\lambda \simeq 1.09\times 10^{25}$, $\phi_0/\Mp\simeq 77.3989$ and
  $A_+/\Mp^3\simeq 2.368 \times 10^{-14}$ and $A_-=0.92A_+$. To make
  the comparison easier we have considered $\nS=1$ for the Planck best
  fit while it is in fact $\nS\simeq 0.96$. Notice that working with
  non-negligible values of $\epsilon_1$ and/or $\epsilon_2$ in the DBI
  Starobinsky model would lead to a significant tilt of the spectrum.}
\label{fig:specplanck}
\end{figure*}

The next step would obviously be to compare in detail the
DBI-Starobinsky model to CMB data, and particularly the recently
released Planck data. This would require interfacing the numerical
code used in this paper to calculate the power spectrum to a CMB code
(typically the \CAMB~code \cite{Lewis:1999bs}), and then in turn to a
code allowing us to explore the corresponding parameter space
(typically the \MCMC~code \cite{Lewis:2002ah}). Moreover, we would
also need to include in the analysis the constraints coming from the
higher correlation functions (see below). Clearly, this is beyond the
scope of the present article since, here, we mainly focus on the
physical properties of the system rather than on data analysis. It is
interesting, however, to have a broad idea about the physical values
of the parameters. To this aim, we have represented in
Fig.~\ref{fig:specplanck} the Planck ``step model'' best fit (solid
blue line) \cite{Ade:2013rta} with the DBI-Starobinsky power spectrum
(solid red line) with $\gamma_\uin\simeq 5$ and $A_-\simeq 0.92
A_+$. Notice that we are now, therefore, in a very different regime to
that discussed above since $A_+$ is similar to $A_-$ and $\gamma $ is
not very large. In this plot, the overall normalisation is arbitrary
(we have normalised the two spectra differently on purpose in order to
make easier the comparison of the two shapes). Notice that the
$x$-axis is now the physical $k$ today and, hence, the position of the
first feature is essentially arbitrary. In particular, it will depend
on the post-inflationary evolution. On the other hand,
Fig.~\ref{fig:specplanck} also indicates that the shape of the
DBI-power spectrum is not the same as the CS-Starobinsky model and, as
a consequence, a rigorous Bayesian exploration of the parameter space
of this model seems to be required before one can conclude whether the
model is ruled out or, on the contrary, whether it could explain the
Planck anomalies and, therefore, improve the fit. In any case, from
the above discussion, it is clear that the ratio $A_+/A_-$ cannot be
too different from one otherwise the amplitude of the oscillations
would obviously be too large (compare for instance Figs.~\ref{fig3}
and~\ref{fig:specplanck}). In addition, $\gamma $ should not be too
large. This has important implications for the exploration of the
parameter space since it means that the relevant regime is in fact the
``non-perturbative'' one \ie the one in which the analytical
expressions derived above are no longer applicable. In this situation,
only our mode by mode code can be used to explore the compatibility of
the model with the data.

Of course, another interesting aspect of the model is to which extent
it produces non-Gaussianities. As is well-known, for slow-roll single
field inflation with a standard kinetic term, the level of
non-Gaussianity is very small, of the order of the slow-roll
parameters, see
Refs.~\cite{Gangui:1993tt,Gangui:1994yr,Wang:1999vf,Gangui:1999vg,Gangui:2000gf,Gangui:2002qc,Maldacena:2002vr}. In
the case of the DBI-Starobinsky model, this is obviously no longer
true. An interesting feature of the model is that, a priori,
non-Gaussianities arise not only from one term, as is usually the case
for non-slow-roll models, but from two (or even various) origins: the
fact that the kinetic term is non standard and the discontinuity of
the derivative of $V(\phi)$ are the main sources of non-Gaussianity in
this scenario. Concretely, the third order action reads
\cite{Seery:2005wm,Chen:2006nt}
\begin{eqnarray}
  S_3 &=& \Mp^2\int {\rm d}\eta \, {\rm d}{\bm x}\biggl[
  -\frac{2a}{3H\cs^2}\left(-\frac{\epsilon_1\delta_1}{3\cs^2\epsilon_X}+u\right)
\zeta'{}^3+\frac{a^2\epsilon_1}{\cs^2}
\left(\frac{\epsilon_1}{\cs^2}+3u\right)\zeta\zeta'{}^2
+\frac{a^2}{2}\frac{\epsilon_1}{\cs^2}\left(\frac{\epsilon_2}{\cs^2}\right)'
\zeta ^2\zeta' \nonumber  \\ & & 
+\frac{\epsilon_1}{2}\zeta \delta^{ip}\delta^{jq}
\partial_i\partial_j\chi \partial_p\partial_q\chi
-2a\frac{\epsilon_1}{\cs^2}\zeta'\delta^{ip}\partial_i\zeta
\partial_p\chi-\frac{a^2}{2}\frac{\epsilon_1^3}{\cs^2}\zeta \zeta '{}^2
+\frac{a^2\epsilon_1}{\cs^2}
\left(\epsilon_1+2\delta_1-u\cs^2\right)
\delta^{ij}\partial_i\zeta \partial_j \zeta \biggr],
\end{eqnarray}
with $\chi\equiv \partial^{-2}\left[a\Sigma
  \zeta'/\left(H^2\Mp^2\right)\right]$, $\Sigma \equiv
\epsilon_1H^2\Mp^2/\cs^2$ and $\epsilon_X\equiv
-\dot{X}/H^2\left(\partial H/\partial X\right)$. The parameter $u$ is
defined by $u=1-1/\cs^2$. In slow-roll canonical inflation, the first
vertex is absent because $u=\delta_1=0$. In DBI inflation, this is no
longer the case and it gives rise to non-vanishing non-Gaussianities
with $\fnl^{\rm eq}\simeq 35u/108$
\cite{Seery:2005wm,Chen:2006nt}. Usually, as already mentioned above,
the other contributions are negligible. On the other hand, if the
kinetic term is minimal and the potential derivative has a
discontinuity, then the third vertex proportional to $\epsilon_2'$ is
the dominant one. Thus an important new feature of the scenario
studied in this paper is that these two vertices are
present. Moreover, one can expect the third vertex to be enhanced by
the factor $1/\cs^2$ since $\cs\ll 1$. The same mechanism should be
valid for the other vertices as well. We are therefore in a rather
complicated set-up in which non-Gaussianities are not easy to
calculate since all terms contribute and since the mode function has a
complicated behaviour. Despite that, it seems clear that the $\fnl$'s
parameters will be quite large, especially compared to the Planck
constraints: $\fnl^{\rm loc}=2.7\pm 5.8$, $\fnl^{\rm eq}=-42\pm 75$
and $\fnl^{\rm ortho}=-25\pm 39$ \cite{Ade:2013ydc}. For instance,
using the DBI equation, one obtains $\gamma \lesssim 12$. But,
clearly, in our case, the constraint should be much tighter since
other terms will contribute. In addition our $\gamma $ is a
time-dependent quantity so the calculation of the contribution of the
first term will be modified. To estimate quantitatively the value of
$\fnl$ is a question that should be addressed by means of numerical
calculations, or maybe using the formalism recently developed in
\cite{Adshead:2013zfa}. It does not come as a surprise since we have
shown before that already the two-point correlation function is an
object difficult to calculate. We conclude that non-Gaussianites will
be a very important probe to constrain the DBI-Starobinsky model.

To end this paper, let us indicate the main directions for future
works. Based on the previous considerations, it seems clear that the
most promising direction is the calculation of non-Gaussianities. On
the theoretical side, we have a new situation, not envisaged
before, in which not only one vertex contributes but many and in a
``coupled fashion'', \ie the fact that $\cs \ll 1$ enhancing the
contribution coming from the discontinuity of $V'$. From the
observational point of view, given the Planck result, it is clear that
the corresponding constraints on the parameters of the model will be
very tight. On the other hand, we have seen that the amplitude of the
superimposed oscillations can be large (or, at least, seems larger
than in the CS-model), even if the parameters are relatively close to
standard slow-roll inflation. As a consequence, a priori, it remains
possible that non-negligible superimposed oscillations improve the
fit to the CMB data (especially by matching the Planck anomalies)
while, at the same time, equilateral non-Gaussianities remain within
the observational bounds. We hope to address this issue in more
detail in the near future.

\section{Acknowledgements}

S.~\'Avila acknowledges financial support from FPI-UAM PhD grants and
`la Caixa' postgraduate scholarships. We thank P.~Brax, L.~Sriramkumar
and V.~Vennin for enlightening discussions.

\bibstyle{aps}
\bibliography{bibliostaro-modified-DS}
\end{document}